\documentclass[aps,floats,floatfix,showpacs,preprintnumbers,amssymb,amsmath,prd,twocolumn,superscriptaddress,nofootinbib,nolongbibliography,reprint]{revtex4-2}

\usepackage{verbatim,mathtools,needspace,enumitem,etoolbox,graphicx,physics,microtype,afterpage,xspace,tabularx,lmodern,multirow}
\usepackage{gensymb}
\usepackage{appendix}
\usepackage{tensor}
\usepackage[normalem]{ulem}
\usepackage[dvipsnames]{xcolor}
\usepackage{xr-hyper}
\definecolor{linkcolor}{rgb}{0.0,0.3,0.5}
\usepackage[unicode, colorlinks=true, linkcolor=linkcolor, citecolor=linkcolor, filecolor=linkcolor, urlcolor=linkcolor, linktocpage, breaklinks]{hyperref}
\usepackage[all]{hypcap}
\usepackage[T1]{fontenc}
\usepackage[utf8]{inputenc}
\usepackage[USenglish]{babel}
\usepackage{microtype}
\usepackage[usenames,dvipsnames]{xcolor}
\hypersetup{colorlinks=true,citecolor=magenta,linkcolor=violet,urlcolor=magenta}

\definecolor{romared}{RGB}{142,0,28}
\newcommand{\nn}{\nonumber}

\newcommand{\be}{\begin{equation}}
\newcommand{\ee}{\end{equation}}

\def\be{\begin{equation}}
\def\ee{\end{equation}}
\newcommand{\beq}{\begin{eqnarray}}
\newcommand{\eeq}{\end{eqnarray}}

\usepackage{makecell}
\usepackage{soul}

\newcolumntype{Y}{>{\centering\arraybackslash}X}

\newcommand{\llangle}{\langle\langle}
\newcommand{\rrangle}{\rangle\rangle}

\makeatletter
\newcommand*{\addFileDependency}[1]{
  \typeout{(#1)}
  \@addtofilelist{#1}
  \IfFileExists{#1}{}{\typeout{No file #1.}}
}
\makeatother

\newcommand*{\myexternaldocument}[1]{%
    \externaldocument{#1}%
    \addFileDependency{#1.tex}%
    \addFileDependency{#1.aux}%
}

\myexternaldocument{Supp}

\begin{document}
\title{Excitation of scalar quasi-normal modes from boson clouds}

\author{Enrico Cannizzaro}
\email{enrico.cannizzaro@tecnico.ulisboa.pt}
\affiliation{CENTRA, Departamento de Física, Instituto Superior Técnico – IST, Universidade de Lisboa – UL, Avenida Rovisco Pais 1, 1049–001 Lisboa, Portugal}
\author{Marco Palleschi}
\affiliation{Dipartimento di Fisica, ``Sapienza'' Universit\`a di Roma \& Sezione INFN Roma1, Piazzale Aldo Moro
5, 00185, Roma, Italy}
\author{Laura Sberna}
\affiliation{Nottingham Centre of Gravity \& School of Mathematical Sciences, University of Nottingham, University Park, Nottingham, NG7 2RD, United Kingdom}
\author{Richard Brito}
\affiliation{CENTRA, Departamento de Física, Instituto Superior Técnico – IST, Universidade de Lisboa – UL, Avenida Rovisco Pais 1, 1049–001 Lisboa, Portugal}
\author{Stephen R. Green}
\affiliation{Nottingham Centre of Gravity \& School of Mathematical Sciences, University of Nottingham, University Park, Nottingham, NG7 2RD, United Kingdom}

\date{\today}

\begin{abstract}
Massive scalar fields on black hole backgrounds generally admit two families of modes: quasi-bound states (QBS) and quasinormal modes (QNM). 
We demonstrate the orthogonality between the two mode families with respect to a relativistic product. 
%
We also find that, although the two families appear on different Riemann sheets of the Green's function of massive scalar perturbations, they can be brought to a single sheet with an appropriate redefinition of the frequency variable. In this variable, it is more natural to see how both mode families can be excited by initial data, and to approximate the Green's function with saddle points.
%
Finally, we investigate the QNM emission from boson clouds -- the latter effectively consisting of a single QBS -- driven by the tidal perturbation of a second compact object.
%
We show that while the resonant emission of QNMs is generally suppressed, QNM transitions may be more prominent when the interaction with the perturber is non-resonant, such as in the dynamical capture of unbound objects, and when the perturber transits close to the light ring.
\end{abstract}


\maketitle

\section{Introduction} 
Bosonic waves scattering off rotating black holes (BHs) can extract rotational energy through the superradiant mechanism whenever the mode frequency $\omega$ satisfies $\omega<m \Omega_H$, where $\Omega_H$ is the angular velocity of the event horizon and $m$ the wave's azimuthal number. For ultralight massive fields, this leads to an instability and the formation of long-lived bosonic condensates composed of quasi-bound states (QBSs) that efficiently extract energy and angular momentum from the BH (see~\cite{Brito:2015oca} for a review). Such systems are known to give rise to striking observable signatures, including characteristic imprints on BH spin distributions and long-lived gravitational-wave emission from the condensate, making them powerful probes of ultralight bosons~\cite{Arvanitaki:2009fg, Arvanitaki:2010sy}.
Extreme mass-ratio inspirals (EMRIs), in which a compact object orbits a supermassive BH for a large number of cycles before plunging, offer another especially sensitive probe of these systems. Indeed, since in EMRIs the interaction between the inspiraling secondary and the BH environment acts continuously throughout the inspiral, even weak effects can accumulate secularly and leave detectable imprints in the gravitational-wave signal observed by space-based interferometers~\cite{Kocsis:2011dr,  Yunes:2011ws, Speri:2022upm}.

EMRIs into BHs surrounded by scalar clouds have garnered significant attention in recent literature (see e.g.~\cite{Baumann:2018vus, Baumann:2019ztm, Baumann:2021fkf, Baumann:2022pkl, Tomaselli:2023ysb, Tomaselli:2024bdd, Brito:2023pyl, Duque:2023seg, Dyson:2025dlj}). A consistent framework for studying such systems employs a Newtonian approximation which reformulates the field equations into a Schrödinger, hydrogen atom-like form, thereby enabling the application of the standard methods of quantum mechanical perturbation theory. Using this approach, Refs.~\cite{Baumann:2018vus, Baumann:2019ztm, Baumann:2021fkf, Baumann:2022pkl, Tomaselli:2023ysb, Tomaselli:2024bdd} showed that the tidal potential of the companion induces resonances between different energy levels of the cloud, and can cause transitions to unbound states akin to ionization in atomic physics. In the process, the companion exchanges energy with the cloud, leading to backreaction on the orbit with observable imprints in the emitted GW~\cite{Cole:2022yzw,DellaMonica:2025zby}.

More recently, these systems have been studied using black-hole perturbation theory~\cite{Brito:2023pyl, Duque:2023seg, Dyson:2025dlj,Li:2025ffh}. This approach captures the key features of EMRIs in boson clouds, previously identified using Newtonian approximations, and generalizes them to the relativistic regime, which is essential for accurately modeling the highly relativistic end of the inspiral. Furthermore, Ref.~\cite{Cannizzaro:2023jle} introduced a relativistic product between superradiant modes, which allowed a relativistic extension of quantum mechanical perturbation theory, and led to the accurate calculation of the frequency shifts due to the self-gravity of the cloud~\cite{Cannizzaro:2023jle}.

So far, studies on EMRIs in boson clouds focused on the observational impact of resonances between different QBSs and transitions to unbound states. QBSs form a discrete spectrum of modes with complex frequencies $\omega_{\rm{QBS}}$ satisfying $\text{Re}(\omega_{\rm{QBS}})^2<\mu^2$, where $\mu$ is the scalar field mass. On the other hand, the unbound spectrum is continuous, composed of real frequencies and can radiate to infinity if $\omega>\mu$.
Bosonic field equations in BH spacetimes admit another family of solutions, called quasi-normal modes (QNMs). QNMs arise in dissipative or open systems -- in this case, modes can radiate to the horizon and to infinity -- and form a complex, discrete spectrum. Unlike QBSs, which only exist for massive fields and are confined in the vicinity of the BH, QNMs can radiate at infinity whenever $\text{Re}(\omega_{\rm{QNM}})^2>\mu^2$.\footnote{In the opposite regime QNMs are evanescent, i.e., their excitation is suppressed and they are irrelevant to observations~\cite{Percival:2020skc, Decanini:2015yba}.} 

While QBSs possess a well-defined limit within the Newtonian approximation -- corresponding to the quantum mechanical hydrogenic bound states -- QNMs are inherently relativistic and do not exist in this regime. Consequently, investigating the role of QNMs in gravitational atoms necessitates a relativistic treatment. 
Like QBSs, scalar QNMs could be resonantly excited in EMRIs. Through the emission of scalar radiation at the horizon and infinity (ionization), this would constitute a novel dissipation channel and affect the EMRI orbital evolution and GW signal. Indeed, a similar phenomenology is known in the gravitational case: gravitational QNMs can be excited during an EMRI by very eccentric or unbound orbits~\cite{Thornburg:2019ukt, Thornburg:2016msc, OSullivan:2014ywd, 10.1143/PTP.72.494}. 

Given the extreme precision required to detect and characterize boson clouds using GW signals, understanding the potential impact of QNM excitation is of significant importance.
The goal of this work is to provide the mathematical framework to assess the excitation of scalar QNMs by scalar clouds and to estimate their influence on the dynamics and observational signatures of EMRIs. To achieve this goal, we adopt the relativistic bilinear form for massive scalar perturbations introduced in Ref.~\cite{Cannizzaro:2023jle} and use it to establish the orthogonality between QBSs and QNMs. We then study the frequency-domain Green’s function of massive scalar fields, where QBSs and QNMs arise as poles on different Riemann sheets, and show how an appropriate redefinition of the frequency variable recasts the problem on a single complex plane, enabling a unified treatment of their excitation.  Finally, we apply time-dependent perturbation theory formulated in terms of the bilinear form to estimate the excitation of scalar QNMs by boson clouds in EMRIs, driven by the tidal perturbation of the secondary compact object.

This work is organized as follows: in Sec.~\ref{sec:Bilinear} we review the product (a bilinear form) for massive scalar modes introduced in~\cite{Cannizzaro:2023jle}. In Sec.~\ref{sec:ortho} we prove, both analytically and numerically, the QBS-QNM orthogonality under this product. In Sec.~\ref{sec:GF} we study the Green Function for massive fields, and show how the QNM-QBS transition can be understood in terms of excitation coefficients. In Sec.~\ref{sec:bosoncloud} we estimate the QNM excitation in an EMRI surrounded by a boson cloud, considering both circular and unbound orbits, describing respectively resonant and non-resonant excitations. Finally, we conclude in Sec.~\ref{sec:conclusions}.

\section{Massive scalar modes around black holes}

\subsection{Quasi-bound states and quasinormal modes}\label{sec:QBSQNM}

The Kerr metric for a BH of mass $M$ and spin parameter $a$ reads, in Boyer-Lindquist coordinates,
\begin{align}\label{eq:BL Kerr met}
{d}s^2 = &-\left(1-\frac{2Mr}{\Sigma}\right) d t^2 - \frac{4Mar\sin^2\theta}{\Sigma} d t d\phi \nonumber \\&+ \frac{\Sigma}{\Delta}d r^2 + \Sigma d\theta^2 + \frac{\Lambda}{\Sigma} \sin^2\theta d\phi^2,
\end{align} 
where $\Delta=r^2+a^2-2Mr$,  $\Sigma=r^2+a^2\cos^2\theta$, $\Lambda = (r^2 + a^2)^2 - \Delta a^2 \sin^2\theta$. We define the event horizon (the greater root $r_\pm$ of $\Delta$) by $r_+$ and the tortoise coordinate $r_*$ as $dr/dr_*=\Delta/\Sigma$. 

The dynamics of a complex, massive scalar field are governed by the Klein-Gordon (KG) equation,
\begin{equation}
    (\Box - \mu^2) \Phi = 0\, .
\end{equation}
Exploiting the symmetries of the Kerr spacetime, we adopt the following ansatz to separate variables,
\begin{equation}\label{eq:multipolar}
    \Phi(t, r, \theta, \phi)= \sum_{\ell m} R_{\ell m}(r) S_{\ell m}(\theta)e^{i m \phi}e^{-i \omega t} \, .
\end{equation}
This ansatz allows to recast the KG equation into a couple of ordinary differential equations for the radial and angular functions,
\begin{align}
    &\frac{ d}{ dr}\Big(\Delta \frac{ d R_{\ell m}}{ dr}\Big)+\Big[\frac{\omega^2(\Delta^2+a^2)^2-4Mam\omega r+m^2 a^2}{\Delta}\\&\nonumber-(\omega^2 a^2+\mu^2r^2+\Lambda_{\ell m})\Big]R_{\ell m}=0  \, ,
\end{align}
\begin{align}
    &\frac{d}{d\theta}\Big(\sin\theta \frac{d S_{\ell m}}{d\theta}\Big)+\Big[a^2(\omega^2-\mu^2)\cos^2\theta-\frac{m^2}{\sin^2\theta}\\&\nonumber+\Lambda_{\ell m}\Big]S_{\ell m}=0 \, ,
\end{align}
where $\Lambda_{\ell m}$ is a separation constant. Solutions to the angular equations are spheroidal harmonics, which reduce to spherical harmonics in the non-rotating limit. The radial equation can instead be solved with suitable boundary conditions at the horizon and infinity. At the horizon, the solution must behave as a purely ingoing wave, while at infinity the leading-order solution reads
\begin{equation}
    \lim_{r\rightarrow\infty}R_{\ell m}\sim r^{-1}r^{(\mu^2-2 \omega^2)/q}e^{q r} \, ,
\end{equation}
where $q=\pm \sqrt{\mu^2-\omega^2}$.
The choice of sign in the exponent determines two distinct families of solutions. With the negative sign, one obtains a family of solutions 
whose radial function decays exponentially at infinity, confining the field near the BH; these correspond to QBSs. Conversely, the positive sign gives non-vanishing, propagative solutions at infinity with $\text{Re} (\omega^2)>\mu^2$, i.e. propagative QNMs, as well as evanescent ones for $\text{Re} (\omega^2)<\mu^2$. 

\subsection{Bilinear form for massive scalars}\label{sec:Bilinear}
Reference~\cite{Green:2022htq} introduced a bilinear form for Weyl gravitational perturbations, and showed that gravitational QNMs with different frequencies are orthogonal under the bilinear form. Subsequently, Ref.~\cite{Cannizzaro:2023jle} extended this product to massive scalar fields and proved the orthogonality of scalar modes with both quasinormal and quasibound asymptotics. In this section, we briefly review this result, and refer to~\cite{Cannizzaro:2023jle} and the review \cite{Berti:2025hly} for additional details.

Starting from the KG equation in Kerr spacetime, one can define the following product: 
\begin{equation}
    \Pi_\Sigma[\Phi_1,\Phi_2] = \int_{\Sigma} d\Sigma_a (\Phi_1 \nabla^a \Phi_2-\Phi_2 \nabla^a \Phi_1),
\end{equation}
where $\Sigma$ is a time-slice with unit normal vector $n^a$. Gauss's theorem allows to verify that, if $\Phi_1, \Phi_2$ are solutions of the KG equations and have compact support on $\Sigma$, the product is unchanged under local deformations of the time slice. Starting from this product, one can build an infinite number of conserved quantities by introducing symmetry operators~\cite{Green:2022htq}. In particular, as the Kerr spacetime is axially-symmetric and stationary, one can consider the $t-\phi$ reflection symmetry operator $\mathcal{J}$, which acting on a scalar field flips sign to the time and azimuthal coordinates. This allows to define the following bilinear form: 
\begin{align}\label{eq:bilinearform_J}
   \langle\langle \Phi_1,\Phi_2 \rangle\rangle 
   =\int\limits_{r_+}^{\infty} dr \int d \Omega  &\bigg[\frac{2 M r a}{\Delta} \left(\mathcal{J}\Phi_{1}\partial_\phi\Phi_2 -\Phi_2\partial_\phi\mathcal{J}\Phi_{1} \right) \nonumber \\
    &+\frac{\Sigma }{\Delta} \left( r^2 + a^2 + \frac{2 M r a^2}{\Sigma}\sin^2 \theta\right) \nonumber\\
    & \times \left( \mathcal{J}\Phi_{1}\partial_t\Phi_2 -\Phi_2\partial_t\mathcal{J}\Phi_{1} \right)
    \bigg] .
\end{align}
One can finally show that the bilinear form is symmetric and the time-translation symmetry operator $L_{t}$ is symmetric with respect to the bilinear form, i.e. $\langle\langle L_{t}\Phi_1,\Phi_2 \rangle\rangle = \langle\langle \Phi_1,L_{t}\Phi_2 \rangle\rangle$. 

As mode solutions in Boyer-Lindquist coordinates diverge both at the horizon (both QNMs and QBSs) and at infinity (QNMs), they do not have compact support at the boundaries.\footnote{To be more precise, only stable QBSs ($\omega_I<0$) diverge as $r_*\to-\infty$, while superradiant modes remain regular. At the threshold $\omega = m\Omega_H$, QBS become bound states with real frequency, and are also regular.} 
In order to apply the bilinear form to mode solutions, Ref.~\cite{Green:2022htq} (see also \cite{Leaver:1986gd,Ma:2024qcv}) introduced a regularization prescription based on the analytic continuation of the integrand in the complex $r-$plane. On the appropriate complex radial contour, the product is well defined on mode solutions, and retains all its properties~\cite{Green:2022htq,Cannizzaro:2023jle}. 

\section{Mode Orthogonality}\label{sec:ortho}
\begin{figure}[t]
\centering
\includegraphics[width=0.9\linewidth]{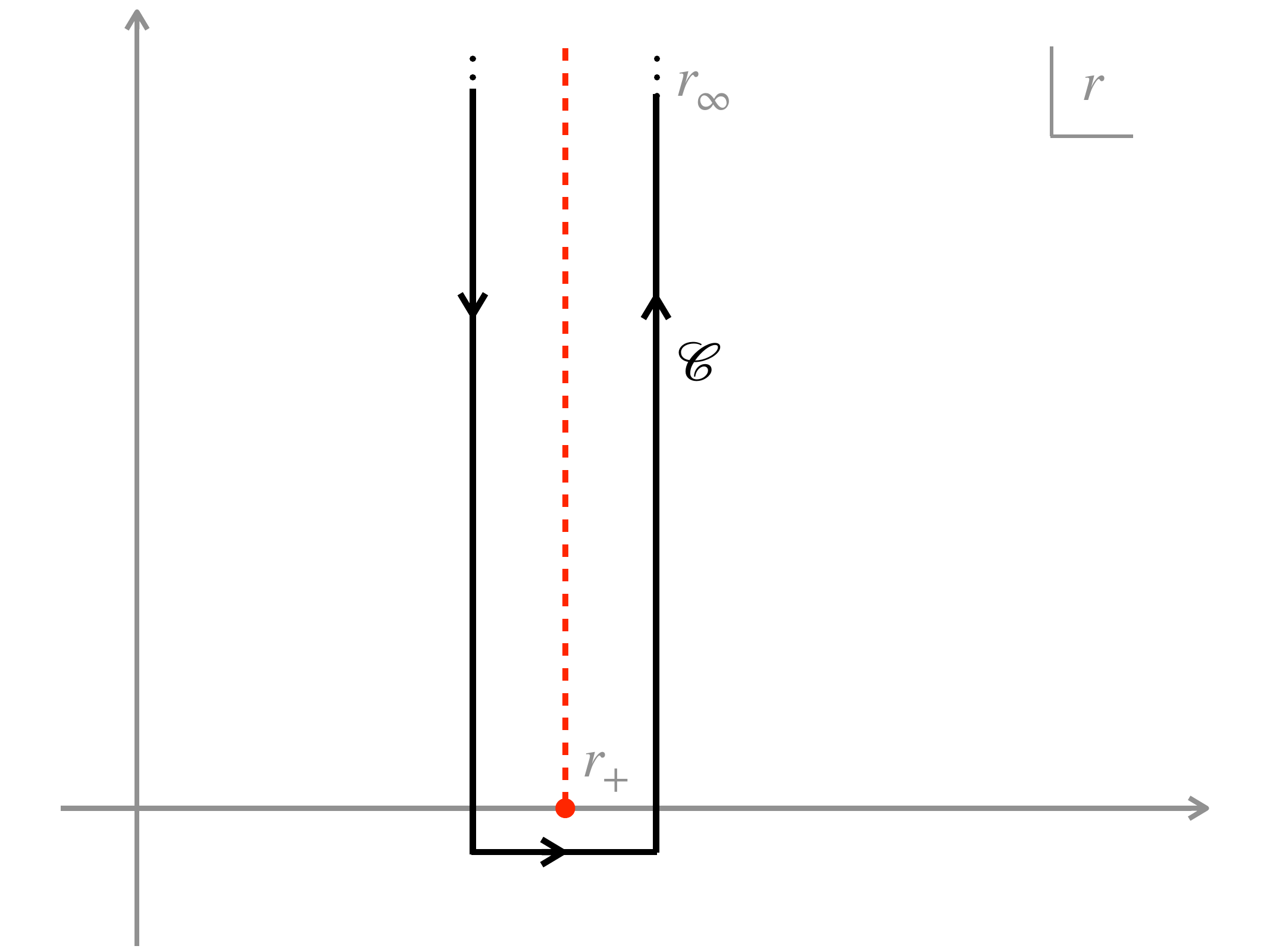}
\caption{Complex contour used to compute the product between a QBS and a QNM with positive real frequencies. We place the branch cut of the mode functions between $r_+$ and $r_+ \to +i \infty$. The contour goes around the branch point. In our numerical implementation in Schwarzschild, the vertical paths are placed at ${\rm Re}(r)= r_+ \pm \epsilon$ with $\epsilon=0.1 M$, while the lower horizontal path is at ${\rm Im}(r)=-\epsilon$. }
\label{fig:contour}
\end{figure}
As shown in~\cite{Green:2022htq, Cannizzaro:2023jle}, the properties of the bilinear form on mode solutions immediately lead to an analytical proof for mode orthogonality. 
Given that the time-dependence of mode solutions is a trivial exponential ($\sim e^{-i \omega t}$), one can exploit the symmetry of the time-translation operator on the bilinear form to obtain trivially,
\begin{equation}
    (\omega_1-\omega_2) \llangle \Phi_1 , \Phi_2 \rrangle=0,
\end{equation}
for a pair of QNMs or QBSs with frequencies $\omega_1$, $\omega_2$. 
Since this statement only assumes wavefunctions to be mode solutions, regular on a suitably defined integration contour, it guarantees not only the orthogonality of QBSs or QNMs, but also the mutual orthogonality between the two families. 
The orthogonality of QNMs and of QBSs individually has also been confirmed numerically in~\cite{Green:2022htq} and~\cite{Cannizzaro:2023jle}, respectively. Here, we provide a numerical demonstration of the orthogonality between the two families of modes.

To do so, in the following we numerically compute the bilinear product between the fundamental $\ell=m=1$ QBS and the fundamental $\ell=m=1$ QNM in Schwarzschild. As previously mentioned, we need to introduce a contour in the complex $r$-plane to ensure the product is finite. We will use the contour described in Refs. ~\cite{Leaver:1986gd, Ma:2024qcv,Berti:2025hly} and shown explicitly in Fig.~\ref{fig:contour}. The branch cut of the QNM and QBS mode functions, originating from the horizon radius $r_+$, is placed parallel to the imaginary axis. The contour goes around the branch point and extends up to $\rm{Im}(r) \rightarrow +\infty$ on both sides of the branch cut. On the contour, the mode decay as $e^{i \omega r} \sim e^{- {\rm Re}(\omega)\rm{Im}(r)}$ as $\rm{Im}(r)\rightarrow +\infty$.\footnote{Here we choose two modes with ${\rm Re}(\omega)>0$; when the product involves one or two negative-frequency (or mirror) modes such that ${\rm Re}(\omega_1+\omega_2)<0$, the branch cut and contour must be reflected across the real axis compared to Fig.~\ref{fig:contour}.} 

In our numerical implementation, we integrate along the vertical paths up to a large value of the radius imaginary part, ${\rm Im}(r_{\infty})$, which ensures numerical convergence. We compute the mode solutions using Leaver's method \cite{Leaver:1985ax,Dolan:2007mj} and integrate numerically using \textsc{Mathematica}. 

\begin{figure*}[ht]
  \centering
  \includegraphics[trim={0.0cm 0.07cm 0.05cm 0.0cm},clip,width=0.33\textwidth]{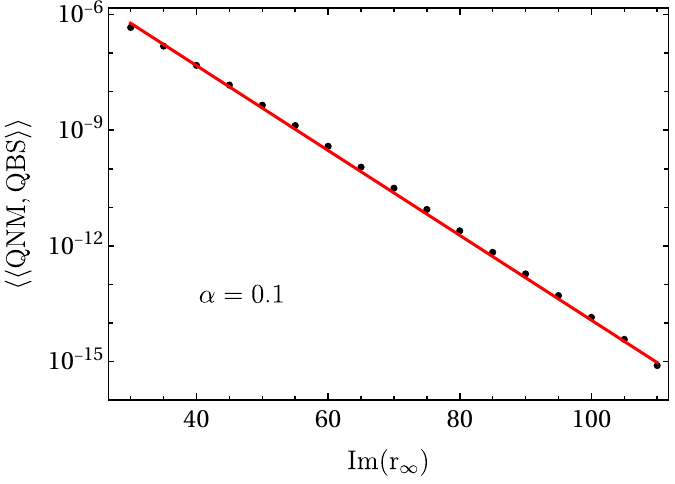}
  \includegraphics[trim={0.6cm 0.07cm 0.05cm 0.0cm},clip,width=0.31\textwidth]{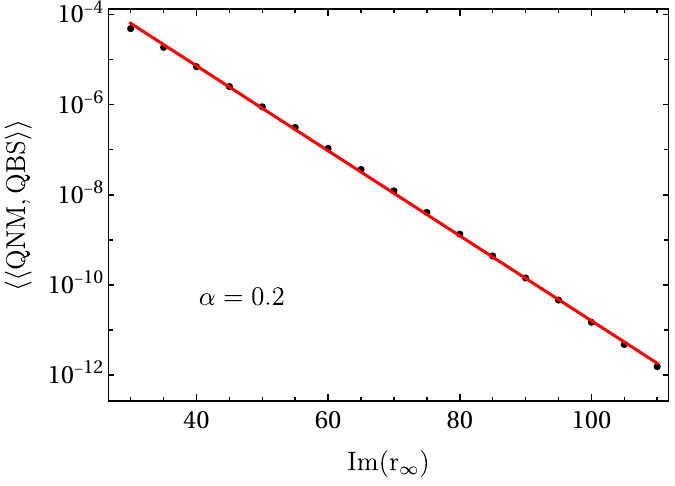}
  \includegraphics[trim={0.6cm 0.07cm 0.05cm 0.0cm},clip,width=0.31\textwidth]{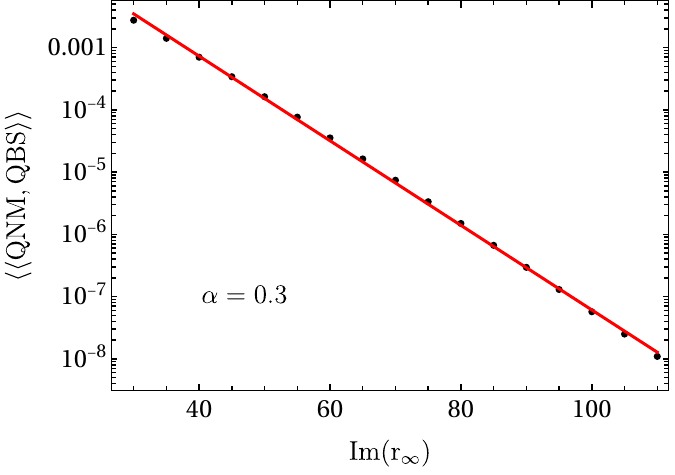}
  \caption{Product between the fundamental $\ell=m=1$ QNM and the fundamental $\ell=m=1$ QBS as a function of the integral cut-off value $\rm{Im}(r_\infty)$. The red curves are exponential fits, showing convergence to zero. Modes are normalized such that $\llangle {\rm QNM, QNM} \rrangle=\llangle {\rm QBS, QBS} \rrangle=1$.}
  \label{fig:orthogonality}
\end{figure*}

Fig.~\ref{fig:orthogonality} shows the relativistic product between a QBS and a QNM as a function of the cut-off value of the complex path ${\rm Im}(r_\infty)$ for different values of the scalar field mass, expressed in terms of the gravitational fine-structure constant $\alpha=M \mu$. As shown in the plot, as the integration is extended to larger values on the complex path, the integral converges to zero exponentially, demonstrating the orthogonality between the modes.

%
\section{The Green's Function of massive fields in Schwarzschild}\label{sec:GF}
In this section, we study the Green's function of massive scalar perturbations, and relate the excitation coefficients of QNMs and QBSs to the bilinear form. We also show how the two mode families, the QNMs and QBSs, naturally appear as poles of the Green's function, although \textit{in two different sheets}, see also discussion in Ref.~\cite{Decanini:2015yba}. For simplicity, we will set the background spin to zero and solve the KG equation in Schwarzschild. Furthermore, we focus on the region of parameter space where QNMs are all propagative, $\text{Re} (\omega^2)>\mu^2$. 

We use the multipolar expansion of Eq.~\eqref{eq:multipolar}, with the spin weighted $s=0$ spheroidal harmonics reducing to the standard spherical harmonics in Schwarzschild. We use a multipolar expansion similar to Eq.~\eqref{eq:multipolar} but with $R_{\ell m}(r)e^{-i\omega t} \to X_{\ell m}(t,r)/r$. The equation for $X$ reads
\begin{equation}
    -\frac{\partial^2}{\partial t^2}X_{\ell m}+\frac{\partial^2}{\partial r_*}X_{\ell m}-V(r)X_{\ell m}=0  \, ,
\end{equation}
for each $l,m$, 
with the potential $V(r)=f(r)(\ell(\ell+1)/r^2+2M/r^3+\mu^2)$.  Owing to the spherical symmetry of the background metric, modes with different multipole numbers $l,m$ are decoupled. Following \cite{Nollert:1992ifk}, we study the initial value problem by performing a Laplace transform,\footnote{Following previous work \cite{PhysRevD.51.353, Berti:2006wq, Glampedakis:2001js}, rather than the standard Laplace parameter $s$, we define $s=i \omega$.}
\begin{equation}
\label{eq:laplacet}
    \hat{X}(\omega,r)=\int_0^\infty e^{i \omega t}X(t,r)dt \, ,
\end{equation}
with the inverse transform defined by
\begin{equation}
\label{eq:laplaceantit}
    X(t,r)=\int_{-\infty+ic}^{\infty+ic} e^{-i \omega t}\Hat{X}(\omega,r)d\omega \, ,
\end{equation}
where $c>0$, i.e. the complex countour of integration is taken above the real axis. Using the Laplace transform, we obtain an inhomogeneous differential equation in the Laplace $\omega$ domain
\begin{equation}
   \omega^2 \hat{X}_{\ell m}+ \frac{d^2}{dr_*}\hat{X}_{\ell m}- V(r)\hat{X}_{\ell m}= I(\omega,r)  \, ,
\end{equation}
where $I(\omega,r)=i \omega X^{(0)}(r)-\dot{X}^{(0)}(r)$ and $X^{(0)}(r)=X(t,r)|_{t=t_0}$ is the initial data. We can solve this inhomogeneous equation via the standard Green's function method. To build the system's Green's function, we introduce two independent solutions of the homogeneous equation, $\hat{X}_{r+}$ and $\hat{X}_{\infty}$. 
Then, the Green's function reads
\begin{equation}
\label{eq:Greenfunction}
    G(\omega, r_*, r_*')= \frac{1}{W(\omega)}\begin{cases}
    \hat{X}_{r+}(\omega, r_*')\hat{X}_{\infty}(\omega, r_*) \ \ \text{if} \ \  r_*'<r_* \\
    \hat{X}_{r+}(\omega, r_*)\hat{X}_{\infty}(\omega, r_*') \ \ \text{if} \ \  r_*'>r_*   \, ,
    \end{cases}
\end{equation}
where $W(\omega)$ is the Wronskian between the two solutions. The full solution of the problem reads, in the frequency domain, 
\begin{equation}
    \hat{X}(\omega, r_*)=\int_{-\infty}^{\infty} G(\omega, r_*, r_*')I(\omega, r_*') dr_*'   \, .
\end{equation}
Finally, the solution in the time domain follows by inserting this in Eq.~\eqref{eq:laplaceantit} and by performing the inverse transform. 

\begin{figure}
{\centering
\includegraphics[trim={8.5cm 8.cm 8.cm 8.cm},clip,width=0.98\linewidth]{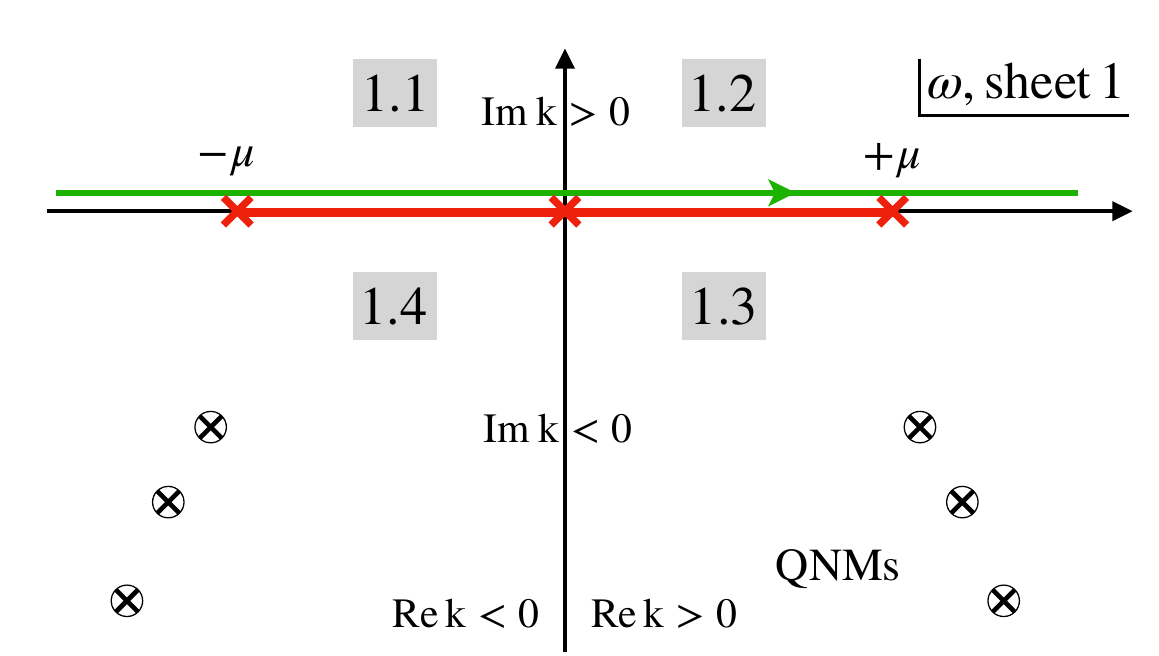}\\
\includegraphics[trim={8.5cm 8.cm 8.cm 7.5cm},clip,width=0.98\linewidth]{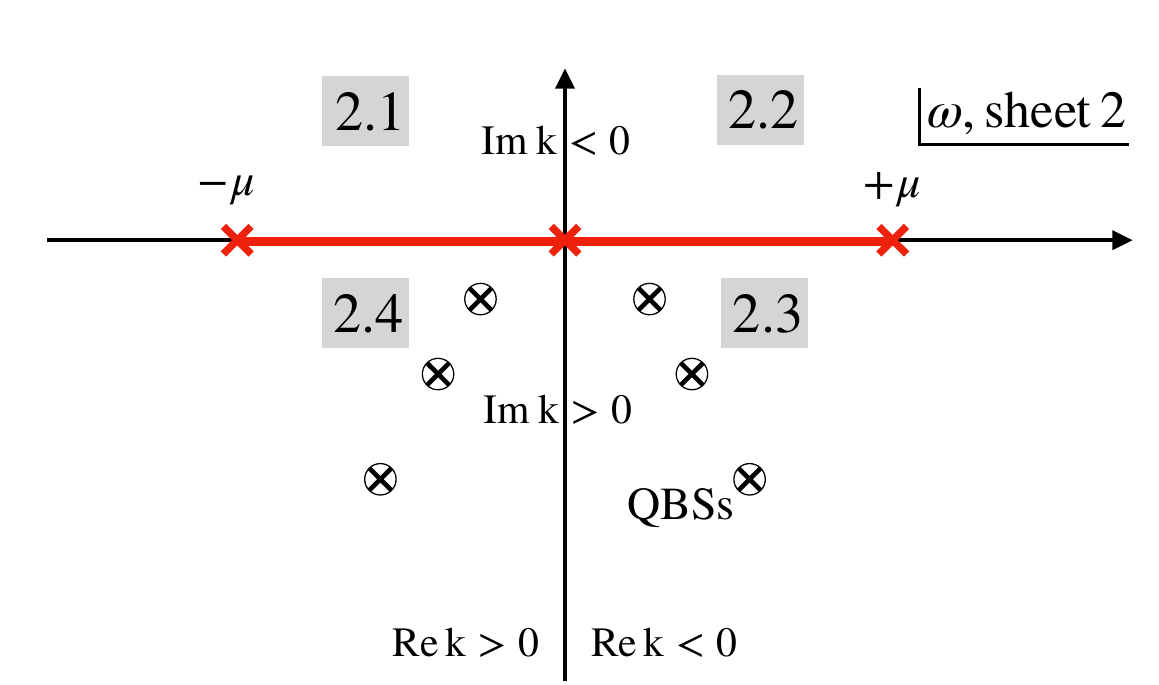}}
\caption{Properties of the Green's function of a massive scalar in the complex plane of the Laplace variable $\omega$. We show the two sheets of the Green's function, arising from the branch cut of the square root $k(\omega)=(\omega^2-\mu^2)^{1/2}$ (red crosses: branch points; our choice for the branch cut: red line). Propagative QNM poles (black crossed circles) lie in the first sheet and have ${\rm Re}(\omega^2)>\mu^2$ (outside the dashed black line), while QBS poles lie in the second sheet and have ${\rm Re}(\omega^2)<\mu^2$. The integration contour runs along and above the real axis in the first sheet.}
\label{fig:contour_omega_sheets}
\end{figure}

We choose the first solution, $\hat{X}_{r_+}$, to behave as a purely ingoing wave at the horizon, while at infinity it is a generic superposition of damped and outgoing waves. Defining the wavevector $k(\omega)=(\omega^2-\mu^2)^{1/2}$, this reads
\begin{align}\label{eq:ABsolution}
    \hat{X}_{r_+} \sim \begin{cases} 
    e^{-i \omega r_*} ,  &r_* \rightarrow -\infty \\
A_\infty(\omega) e^{i k(\omega) r_*}+B_\infty(\omega) e^{-i k(\omega) r_*} , &r_* \rightarrow \infty   \, .
\end{cases}
\end{align}
%
Note that at leading order the behavior at the horizon is the same for massive and massless fields, as the mass correction is subdominant in this limit. 
As for the second solution $\hat{X}_{\infty}$, we define it in the Riemann sheet of $k(\omega)=(\omega^2-\mu^2)^{1/2}$ where $\rm{Im}(k)<0$ for $\rm{Im}(\omega)<0$ and $\rm{Re}(k)>0$ when $\rm{Re}(\omega)>0$ (sheet 1, see Fig.~\ref{fig:contour_omega_sheets}) as
\begin{equation}\label{eq:infsolution}
    \hat{X}_{\infty}\sim e^{i k(\omega) r_*}
    ,  \ \ r_* \rightarrow \infty   \, ,
\end{equation}
corresponding to an outgoing wave at infinity. Note that these solutions can be defined by analytic continuation on the two Riemann sheets associated with the square root in $k$.

To perform the inverse Laplace transform \eqref{eq:laplaceantit}, we study the analyticity properties of the Green's function in the complex $\omega$ plane, see Fig.~\ref{fig:contour_omega_sheets}. The Green's function inherits from the two solutions \eqref{eq:ABsolution} and \eqref{eq:infsolution} two branch points, at $\omega=\pm \mu$. Our choice is to place the branch cut between the two points, along the real axis. As a result, the problem admits two sheets, corresponding to the two signs of the square root in the exponential terms $e^{\pm i k r}$~\cite{Decanini:2015yba}. We have defined the Green's function in the first sheet, corresponding to an outgoing boundary condition at infinity. In this sheet, the Green's function has poles (corresponding to zeros of the Wronskian) in the region $|\omega|>\mu$, which can only correspond to QNMs: they are outgoing ($\rm{Re}(k)>0$ when $\rm{Re}(\omega)>0$) and they are not spatially confined ($\rm{Im}(k)<0$). In the second sheet, the Green's function has poles in the region $|\omega|<\mu$ corresponding to QBSs, since $k$ has a positive imaginary part and \eqref{eq:infsolution} becomes a damped solution at infinity. 
The Green's function has an additional branch point at $\omega=0$, but the corresponding branch cut does not affect its mode solutions~\cite{Leaver:1986gd, Shahar1998, Koyama:2001ee, Koyama:2001qw}. 

It is often useful to re-express the integral over $\omega$ in the inverse Laplace transform in terms of a closed integral, to make the contribution of the poles and branch cuts more explicit, see e.g.~\cite{Leaver:1986gd, Nollert:1992ifk}. In the case of massive perturbations, this standard approach is complicated by the existence of poles on two distinct sheets. In this case, we cannot use the standard contour closed in the lower half $\omega$ plane to express the solution in terms of both QBSs and QNMs, and to compute their excitation coefficients. 
In order to make both of these contributions explicit, we take a different approach and follow a method first introduced in optics for leaky wave-guides~\cite{Leaky1, 10.1121/1.4973313}. These systems have a modal structure similar to that of massive fields around BHs, with two  distinct families of modes lying on two different sheets of the same Riemann plane.

\begin{figure}
\centering
\includegraphics[trim={11.cm 7.5cm 11.cm 7.5cm},clip,width=0.95\linewidth]{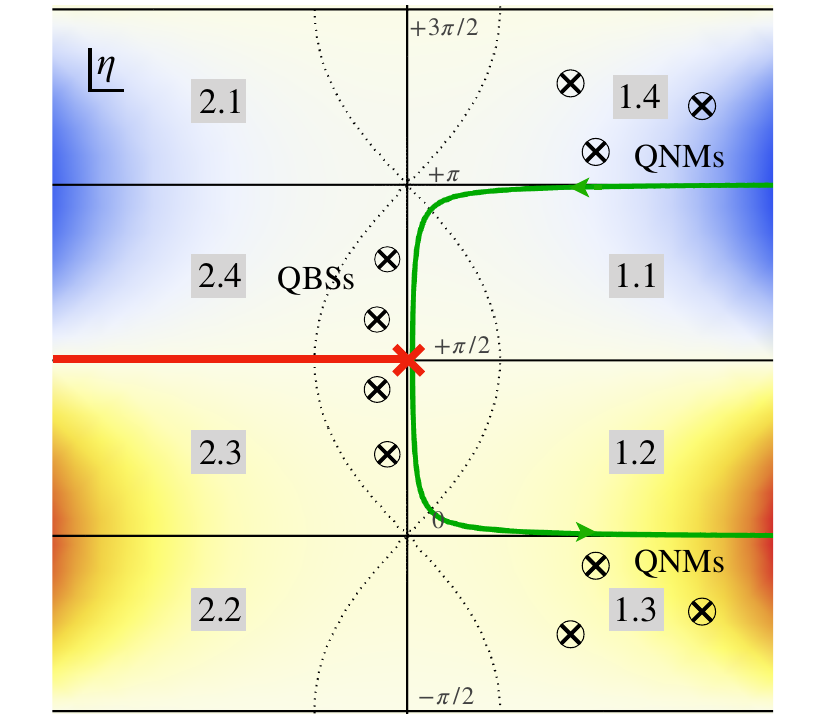}
\caption{Properties of the Green's function in the new complex variable $\eta$, see Eq.~\eqref{eq:eta}. We follow the same conventions as in Fig.~\ref{fig:contour_omega_sheets}: the solid green line is the original integration contour; the dotted black line marks $|\omega|=\mu$; black crossed circles mark the poles; the red cross marks the Green's function's remaining branch point at $\omega=0$; black lines denote the old axis ${\rm Re}\,\omega=0$ and ${\rm Im}\,\omega=0$. The background color corresponds to the sign of the real part of $\omega$ (red for positive, blue for negative). 
}
\label{fig:contour_other}
\end{figure}

We define a new complex variable, $\eta$, that maps the two-sheeted $\omega$-plane into a strip of the $\eta$-plane,
\begin{equation}\label{eq:eta}
    \omega = \mu \cosh{\eta}, \quad k = \mu \sinh{\eta}\, .
\end{equation}
In order to represent the original Riemann surface, we restrict the new variable to the strip $-\pi/2<{\rm Im}(\eta)\le3 \pi/2$. Figure~\ref{fig:contour_other} shows how all regions and features of the $\omega$-plane (Fig.~\ref{fig:contour_omega_sheets}) map onto the new plane. Each of the eight quadrants of the two $\omega$-sheets is now mapped into a semi-infinite strip in the $\eta$ plane, and thus the new complex plane encompasses both QNMs and QBSs.

The variable $\eta$ is also used to identify a convenient representation of the Green's function, in terms of saddle points and steepest descent contours \cite{Leaky1}. This can be seen analytically, e.g., in the limit $r_*\to \infty$ and $r'_*\to -\infty$. In this limit, the integrand of the inverse Laplace transform reads
\begin{equation}
    e^{-i\omega t} \hat{X}(\omega, r) \sim f(\eta)e^{\mu\Psi(\eta)} \, ,
\end{equation}
with
\begin{equation}
    \Psi(\eta) = - i \left( (t+r_*')\cosh\eta  - r_* \sinh \eta \right) \, .
\end{equation}
and $f$ a function slower than exponential.

The Green's function admits saddle points where $\Psi^\prime(\eta_{\rm sp})=0$ and associated steepest-descent contours in the complex $\eta$ plane. 
The saddle points are located at
\begin{equation}\label{eq:saddlepoints}
    \eta_{\rm sp,j} = \log\left(-i\frac{r_*+t+r'_*}{r_*-t-r'_*}\right) + i \pi j \, .
\end{equation}
Along these contours, the integrand is purely decaying: ${\rm Re} \Psi(\eta)<0$, ${\rm Im} \Psi(\eta)={\rm Im} \Psi(\eta_{\rm sp})$. As a result, the integral in $\eta$ performed along the steepest-descent contours is absolutely convergent, a convenient choice to deform the original contour. This heuristic is formalized in the mathematical Picard-Lefschetz theory, described for example in \cite{Feldbrugge:2017kzv}. 

\begin{figure}
\centering
\includegraphics[trim={11.cm 7.5cm 11.cm 7.5cm},clip,width=0.9\linewidth]{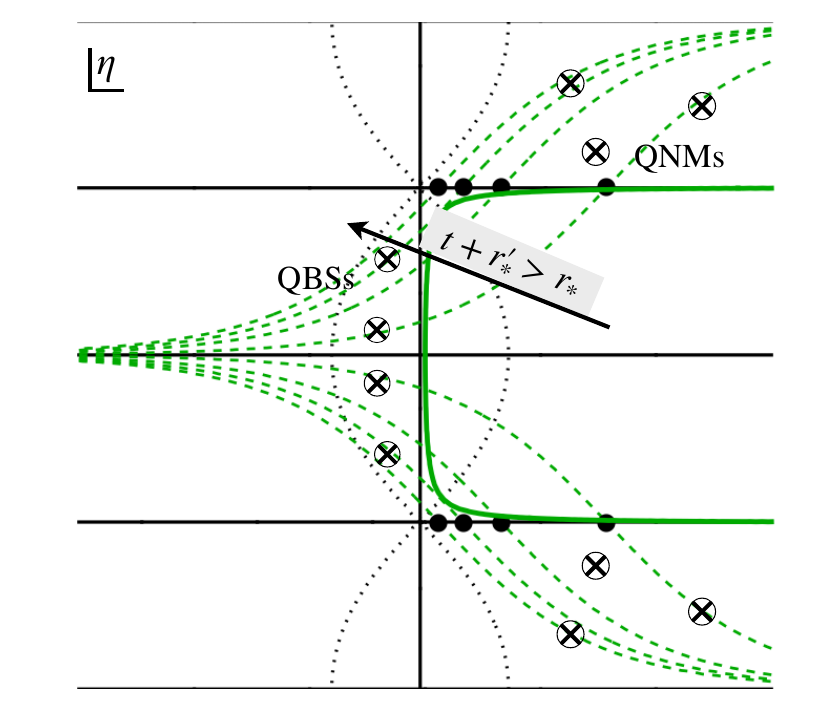}
\caption{Location of the saddle points (black dots) and steepest-descent contours (dashed green lines) of the massive scalar Green's function in Schwarzschild in the complex $\eta$ plane, at large $t+r'_*>r_*$, with $r_*$ large and $r'_*$ approaching the horizon. See previous Fig.~\ref{fig:contour_other} for conventions.}
\label{fig:contour_time}
\end{figure}

Figure~\ref{fig:contour_time} shows the location of the relevant saddle points and relevant contours onto which to deform the original integration path. The figure also shows that, when deforming the original contour onto the steepest-descent ones, the Green's function inevitably picks up the contribution of both QBS and QNM poles. 
Note that the location of the saddle points and steepest-descent contours depends on $t$, and $r_*$ and $r'_*$. At late times, the steepest-descent contours will tend to encompass more and more modes (poles), see Fig.~\ref{fig:contour_time}. 
%

Focusing only on the contribution of the poles and using the residue theorem in a standard fashion, we can write the solution in the time domain at late times and at large $r$ as
\begin{align}
\label{eq:Laplacesol}
    X(t,r) \sim &\sum_n C^{\rm QBS}_n e^{-i \omega_n t}e^{-\sqrt{\mu^2-\omega_n^2} r} \nn \\
     &+ \sum_n C^{\rm QNM}_n e^{-i \omega_n t}e^{+\sqrt{\mu^2-\omega_n^2} r} \, .
\end{align}
Here, from the residues of the QNM and QBS poles, we have defined the QNM and QBS excitation coefficients, as in the case of gravitational QNMs~\cite{Berti:2006wq}, 
\begin{align}
\label{eq:excitcoeff}
    C^{ K}_n=  \frac{iA^{ K}_{\infty}(\omega_n)}{\left. d W^{ K}/d\omega \right|_{\omega_n}} \int_{-\infty}^{\infty} dr_*' \hat{X}^{ K}_{\omega_n}(\omega_n, r_*')I(\omega_n, r_*')   \, , 
\end{align}
where $K=$QNM, QBS indicates the asymptotic behavior of the mode solutions as $r_*\to +\infty$, and $A_\infty^K$ is the asymptotic amplitude defined in Eq.~\eqref{eq:ABsolution}.

The integral appearing in the excitation coefficients represents the overlap between the initial data and the modes. In Ref.~\cite{Green:2022htq}, it was shown that the QNM excitation coefficients can be written in terms of the bilinear form-projection of the initial data onto the (normalized) modes. The same argument applies to the massive case, and in particular to QBSs. We can therefore write  
\begin{align}
\label{eq:excitcoeff_product}
    C^{ K}_n=   \frac{\langle\langle X^K_n, X^{(0)} \rangle\rangle}{\langle\langle X^K_n, X^K_n \rangle\rangle}  \, .
\end{align}
%
As a consequence of Eq.~\eqref{eq:excitcoeff_product} and mode orthogonality, an initial state populated of QBSs would not spontaneously excite QNMs, and viceversa, at linear order in the perturbation.

\section{QNM excitations in EMRIs surrounded by boson clouds}\label{sec:bosoncloud}
In previous sections, we showed how QNMs and QBSs are mutually orthogonal, and how both can be simultaneously excited by initial data. In this section we will provide a phenomenological application of our framework and show how transitions between the two mode families can be induced by an external source, and estimate the impact of such transitions. In particular, we will focus on transitions sourced by the tidal potential generated by a secondary in an EMRI. 

For simplicity, we assume the primary BH is non-spinning. Strictly speaking, boson clouds generated via superradiance require a Kerr background, since superradiant amplification relies on the BH’s rotation. However, the essential features of the transitions we study -- namely, the coupling between QNMs and QBSs and their response to an external tidal source -- can already be captured in the nonrotating limit. A generalization to the Kerr case will be carried out in future work.

We consider a BH with mass $M$ surrounded by a boson cloud, perturbed by a small companion of mass $M_*$. The tidal potential generated by a companion reads, in the Newtonian approximation \cite{Baumann:2019ztm, Tomaselli:2023ysb},
\begin{equation}
V(t, \textbf{r})=-\sum_{\ell_*=0}^\infty \sum_{m_*= -\ell_*}^{\ell_*}\frac{4 \pi q \alpha}{2 \ell_*+1}Y_{\ell_* m_*}(\theta_*, \phi_*)Y^*_{\ell_* m_*}(\theta , \phi)G(r) \, ,
\end{equation}
where $q=M_*/M\ll 1$ is the mass ratio,
\begin{equation}\label{eq:gnewt(r)}
G(r)=
\begin{cases}
\dfrac{r^{\ell_*}}{R^{\ell_*+1}}\;\Theta(R_*-r)
+\dfrac{R_*^{\ell_*}}{r^{\ell_*+1}}\;\Theta(r-R_*)
&\text{for }\ell_*\neq 1,\\[6pt]
\displaystyle\Big(\dfrac{R_*}{r^2}-\dfrac{r}{R_*^2}\Big)\;\Theta(r-R_*)
&\text{for }\ell_*=1.
\end{cases}
\end{equation}
and $(R_*(t), \theta_*(t), \phi_*(t))$ are the coordinates of the companion in a coordinate system centered at the larger BH.

Reference~\cite{Cannizzaro:2023jle} showed, using a Hamiltonian formalism, that an external gravitational potential $\delta V$ sources transitions of the cloud in a state $\Phi_1$ to new states $\Phi_2$ regulated by the relativistic matrix element $\llangle \Phi_1, \delta V \Phi_2 \rrangle$, with the product given in Eq.~\eqref{eq:bilinearform_J}. 
While in principle, within our relativistic framework, one should employ the fully relativistic tidal potential, its explicit construction and analytic treatment are technically demanding. Therefore, following Ref.~\cite{Cannizzaro:2023jle}, we adopt a semi-relativistic approximation in which the modes and overlap are computed relativistically while the interaction is mediated by the Newtonian potential. 

\begin{figure}[t]
\centering
\includegraphics[width=0.95\linewidth]{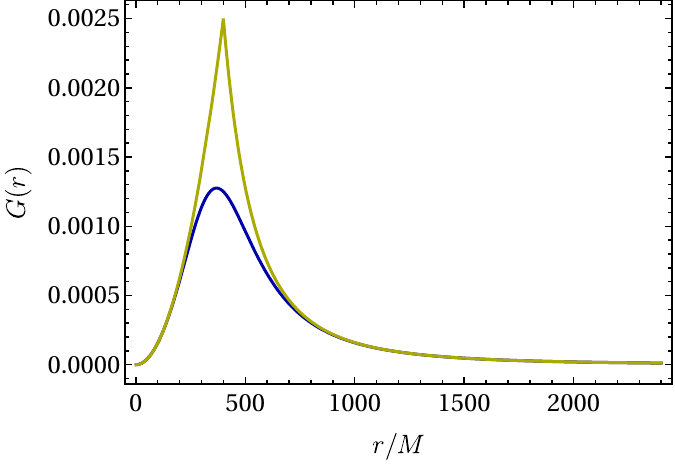}
\caption{Comparison between the Newtonian potential of a point-particle (yellow line) and our toy-model, smooth potential (blue line) for $R_*/M=400$. The toy model accurately reproduces the Newtonian potential in both asymptotic regimes.
}
\label{fig:potential}
\end{figure}

The potential \eqref{eq:gnewt(r)} is continuous but not differentiable, because of the Heaviside step functions. Hence, analytical continuation to the complex plane -- necessary to evaluate the product in the relativistic matrix element (see Sec.~\ref{sec:Bilinear}) -- would not be possible. The discontinuity in the derivative at $r=R_*$ arises because the secondary is modeled as a point particle; including finite--size effects would naturally smooth this out. Therefore, as a toy model we consider a potential which smoothly interpolates between the two regimes and is differentiable at $R_*$, while still closely approximating the original expression away from the particle (as one would expect finite-size effects to do).\footnote{Note that similar smoothing procedures for the point-article potential are commonly adopted in Newtonian physics, see e.g.~\cite{2020ApJ...892...65M, 2024MNRAS.534...39B}.} Hence, our toy-model potential reads
\begin{equation}\label{eq:g(r)}
G(r)=
\begin{cases}
\frac{(r/R_*)^{\ell}}{1 + (r/R_*)^{2\ell+1}} 
&\text{for }\ell_*\neq 1,\\[6pt]
\displaystyle\Big(\dfrac{R_*}{r^2}-\dfrac{r}{R_*^2}\Big)\;\Big(\frac{1}{1+e^{-(r+R_*)/2}}\Big)
&\text{for }\ell_*=1.
\end{cases}
\end{equation}
As an example, a comparison between the two potentials for $\ell_*=2$ is shown in Fig.~\ref{fig:potential}. We tested this approximate potential for different values of $\alpha$ and $R_*$, and found that it introduces at most a $O(10\%)$ error in the matrix element.

\subsection{Overlap integrals}

We are now ready to estimate the QNM excitation induced by an EMRI on a scalar cloud. Note that in the following, we normalize both QB and QN modes such that $\llangle \rm{\Phi}_{lmn},  \rm{\Phi}_{l'm'n'} \rrangle=\delta_{l l'}\delta_{m m'}\delta_{n n'}$. We first identify the allowed transitions and estimate their matrix elements.
 The angular part of the overlap integral allows only the excitation of modes that satisfy the angular selection rules
\begin{align}\label{eq:selection}
    &-m_j+m_i-m_*=0 \, ,\nonumber \\ &
    |\ell_j-\ell_i|\leq \ell_* \leq \ell_i+\ell_j \, ,\\&\nonumber
    \ell_i+\ell_j+\ell_*=2p \ \ \text{for}\ \  p \ \in \mathbb{Z} \, ,
\end{align}
where $i,j$ denote the initial and final state, respectively. As the dominant superradiant cloud is the fundamental $\ell=m=1$, in the following we will always assume this as the initial state. In this case, the most natural transition allowed by the selection rules is between the fundamental $\ell_i=m_i=1$ QBS and the $\ell_j=1, m_j=0, -1, 1$ QNMs, sourced by the quadrupole tidal potential $\ell_*=2$. The fundamental $\ell_j=1$ QNM frequency is weakly-dependent on $\alpha$, and its absolute value is $\approx 0.35/M$. As we require this value to be higher than $\alpha$ for the mode to propagate toward infinity, we will consider values of $\alpha$ slightly below this threshold. 

\begin{figure}[t]
\centering
\includegraphics[width=\linewidth]{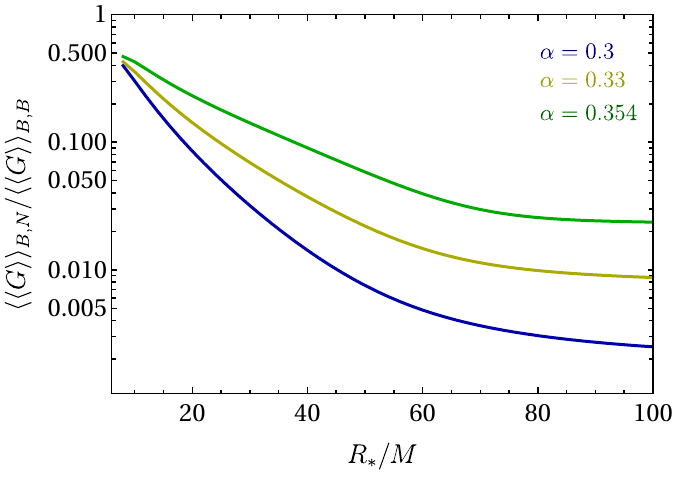}
\caption{Ratio between the fundamentals $l=1$ QN-QB and the QB-QB overlap in the presence of a tidal potential as a function of $R_*$ for different $\alpha$. As $\alpha$ increases and the cloud becomes relativistic, the overlap with QNMs becomes more prominent even when the secondary is further away.}
\label{fig:overlap}
\end{figure}

In this subsection, we compute the QN-QB overlap mediated by a secondary at a fixed radius. In Sec~\ref{sec:capture}. we will generalize this to a dynamical secondary. 
Figure~\ref{fig:overlap} shows the ratio between the QN-QB and the QB-QB overlap integral for different values of $\alpha$ as a function of the companion radius $R_*$. While for large $R_*$ the QN-QB overlap integral is suppressed, it remarkably becomes comparable to the QB-QB one for small values of $R_*$. Furthermore, the QNM is excited more for larger $\alpha$. In this regime, the scalar cloud is more compact and relativistic, lying closer to the BH. As a result, its overlap with the QNMs -- whose wavefunctions peak near the light ring -- is enhanced.  

In principle, resonant transitions with $\ell_j=0$ QN or QB modes, sourced by the dipolar tidal potential $\ell_*=1$, are also possible. However, these transitions are strongly suppressed in the QNM case. The $\ell_j=0$ QNM lies at a frequency of $\approx0.11/M$, so one would require $\alpha$ slightly below this value for this mode to propagate to infinity. In that regime, the cloud is highly non-relativistic and peaked far from the BH, leading to a very weak overlap with the QNM mode function. Indeed, we find that compared to the $\ell_j=1$ case shown in Fig.~\ref{fig:overlap}, the overlap with the QN $\ell_j=0$ mode is suppressed by 5–6 orders of magnitude. 
On the contrary, transitions to $\ell_j=2$ QNMs, mediated by octupolar perturbations $\ell_*=3$, share similar features to the $\ell_j=1$ case, becoming comparable to QB-QB transitions at small $R_*$ and decaying further away. 

The calculation of the overlap is only the first step to estimate the QNM excitation due to the companion. Such transitions can be classified into resonant and non-resonant~\cite{Tomaselli:2023ysb}. A companion on a circular orbit induces a tidal potential with the characteristic frequency of the orbit, which evolves adiabatically during the inspiral as the companion looses energy due to gravitational wave (GW) emission and its interaction with the cloud. This periodic source leads to resonances, with different resonances excited at different times during the inspiral. On the contrary, a particle approaching the BH on an unbound orbit breaks adiabaticity, inducing non-resonant transitions. In the following, we will analyze both regimes, starting from the resonant case. 

\subsubsection{Resonant transitions in circular orbits}
In boson clouds, a companion in circular orbits induces a resonant transition between a QBS with quantum numbers $\ell_i, m_i, n_i$ and a state with quantum numbers $\ell_j, m_j, n_j$ whenever~\cite{Baumann:2018vus,Brito:2023pyl}:
\begin{equation}\label{eq:resonantcondition}
    \Omega_p=\frac{\text{Re}(\omega_{\ell_j, m_j, n_j})- {\rm Re}(\omega_{\ell_i, m_i, n_i})}{m_j - m_i} \, ,
\end{equation}
where $\Omega_p=\sqrt{M/r^3}$ is the orbital frequency. It is immediate to see that this condition in principle allows resonant transitions into QNMs in some regions of the parameter space. For example, transitions to the $\ell=1, m=-1$ QNMs for $\alpha=0.33$ arises whenever the secondary is counter-rotating at an orbital radius $R_* \approx 23 M$. Nevertheless, Fig.~\ref{fig:overlap} immediately shows a limitation of this mechanism: transitions into QNMs are strongly suppressed by the overlap integral unless $R_*$ is very small. In principle, the resonance radius can approach the ISCO when $\alpha$ is chosen such that the numerator of Eq.~\eqref{eq:resonantcondition} is large. 

Even when the resonance condition is satisfied at small orbital radii, where the overlap is enhanced, another limitation arises. Resonant QBS-QNM transitions will also be suppressed by the decay rate of the QNMs. Indeed, the width of resonances between modes of a discrete complex spectrum scales as $\rm{Im(\omega)}^{-1}$~\cite{Brito:2023pyl}. Hence, such resonances are typically prominent in the case of long-lived modes with a small imaginary part, while they are suppressed for the short-lived QNMs. Given these two suppression factors, we expect resonant transitions into QNMs to be subdominant in boson clouds. 

To confirm this statement, we computed the scalar fluxes at infinity of an EMRI embedded in a boson cloud using the relativistic framework introduced in Ref.~\cite{Brito:2023pyl}. We did not find any trace of resonances in the flux at infinity, confirming that resonant excitation of QNMs is completely suppressed. 

One might wonder whether, in some region of the parameter space, the imaginary part of scalar QN frequencies becomes small enough to lead to a significant excitation. For massive scalar fields, QNMs can become long-lived in two distinct regimes. In the large-$\alpha$ limit, their resonant excitation is enhanced, giving rise to so-called giant ringings~\cite{Decanini:2015yba, Decanini:2014bwa}. However, this scenario is of limited observational relevance for boson clouds, since in this regime the scalar mass is so large that the superradiant instability timescale is exponentially suppressed.
A second regime in which long-lived QNMs appear is that of nearly extremal BHs~\cite{Cavalcante:2024kmy}. Yet this case is also observationally disfavored for boson clouds. If a stationary cloud forms around a nearly extremal BH at the end of the superradiant instability, the cloud mass remains negligible: only a minimal amount of spin energy can be extracted, leading to a very small cloud. Thus, this regime is also unlikely to yield observable signatures. 

\subsubsection{Non-resonant transitions in parabolic orbits}\label{sec:capture}

We now focus on non-resonant processes, and estimate the energy loss due to QNM emission of a secondary crossing a boson cloud while following a parabolic orbit. This is relevant in the context of dynamical capture of compact objects by massive BHs in nuclear star clusters and the formation of extreme mass ratio inspirals, see e.g.~\cite{Amaro-Seoane:2012lgq,AmaroSeoane2017}.
Parabolic orbits present two key advantages over circular ones. Firstly, the process is non-resonant, and therefore less affected by the short lifetime of QNMs. Secondly, in a parabolic orbit the particle can pass very close to the BH, significantly enhancing the QN–QB overlap, which is maximal in this regime, as illustrated in Fig.~\ref{fig:overlap}.

Previous studies, using a Newtonian approximation, showed that energy loss from transitions into unbound modes and various QBSs can significantly exceed the loss due to GW emission~\cite{Zhang:2019eid,Tomaselli:2023ysb}. The Newtonian computation can be generalized to the relativistic case. In particular, the treatment of the discrete spectrum -- now including QNMs -- carries over straightforwardly: in both regimes, the scalar wavefunction can be expressed as a discrete sum of modes, and the occupation number of each mode excited by the fly-by particle is obtained by solving a Schrödinger-like equation. In the relativistic setting, the perturbed scalar field due to the tidal potential can be expanded as a combination of modes with time-dependent amplitudes, $\Phi=\sum_n c_n(t) \Phi_n$~\cite{Cannizzaro:2023jle}. As shown in \cite{Cannizzaro:2023jle}, using an Hamiltonian formalism mirroring quantum mechanical perturbation theory, one obtains
\begin{equation}\label{eq:schroedinger}
    \frac{\partial}{\partial t} c_n = \sum_q c_{q}\frac{\llangle F_{n},  \delta H F_{q}\rrangle}{\llangle \Phi_n, \Phi_n \rrangle} \, ,
\end{equation}
where \(F = (\Phi, \Pi)^{T}\) denotes the phase-space field vector, with 
\(\Pi\) the conjugate momentum of \(\Phi\), and the bilinear form in phase space is defined as
\begin{equation}
    \llangle F_1, F_2 \rrangle
    = \int_{\Sigma} \left( \, \Phi_2\,\mathcal{J} \Pi_1 +\Pi_2 \,\mathcal{J}\, \Phi_1   \right)\, \mathrm{d}^3 x \, .
\end{equation}
Finally, $\delta H$ can be rewritten for a gravitational perturbation as $\delta H = \delta V H_0$, $H_0$ being the zeroth order Hamiltonian matrix (see ~\cite{Cannizzaro:2023jle} for further details).
Following~\cite{Tomaselli:2023ysb}, we assume that the initial boson cloud is populated only by the $\ell_i=m_i=1$ QBS and that the perturbation is small enough such that the coefficient of the background state is $c_{\ell_i m_i} \approx 1$ throughout the evolution.
Under this assumption, Eq.~\eqref{eq:schroedinger} can be solved by a simple time integration in the limit $t \rightarrow \infty$, providing a relativistic version of the Newtonian solution of \cite{Tomaselli:2023ysb},
\begin{equation}\label{eq:integral}
    c_n=-i \int_{-\infty}^{\infty}dt\frac{\omega_{\ell_i, m_i} \llangle F_{ n},  \delta V F_{\ell_i, m_i}\rrangle}{\llangle \Phi_n, \Phi_n \rrangle} \, .
\end{equation}
Indeed, assuming $\omega_{\ell_i m_i}\approx \mu$ the Schrodinger-like equation \eqref{eq:schroedinger} coincides with the Newtonian one, with the hydrogenic scalar product replaced by the bilinear form in the Hamiltonian framework (see Eq.~(3.5) in~\cite{Tomaselli:2023ysb}). 

In the following, we include QNMs in the ansatz and use Eq.~\eqref{eq:schroedinger} to compute the QNM excitation coefficient. 
In practice, we insert the overlap obtained with our bilinear form in the numerical routine developed in Ref.~\cite{Tomaselli:2023ysb}, and publicly available on GitHub \cite{GrAB}, which provides the particle's trajectory and performs the time integration.

\begin{figure}[t]
\centering
\includegraphics[width=\linewidth]{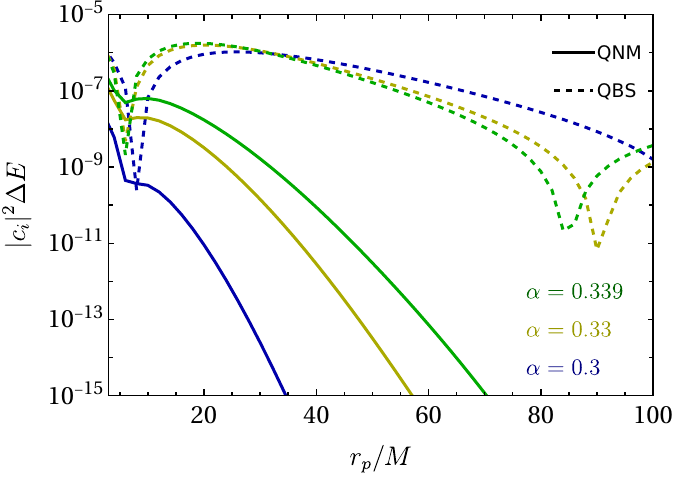}
\caption{Energy loss of a fly-by particle due to non-resonant transitions of a $\ell_i=m_i=1$ QBS into $\ell_j=1$ QNMs (continuous lines) and QBSs (dashed lines) as a function of the periapsis of the orbit. As $\alpha$ increases, the energy loss due to QNM excitation becomes comparable with the QBS one whenever the periapsis is small.}
\label{fig:Elost}
\end{figure}

Following~\cite{Tomaselli:2023ysb}, in Fig.~\ref{fig:Elost} we show the total (summed over all $m_j$) square-amplitude of the $\ell_j = 1$ QN and QB states multiplied by the real frequency difference with the initial $\ell_i = m_i = 1$ cloud -- providing a heuristic estimate of the energy loss -- as a function of the periastron radius of the unbound orbit, for different values of $\alpha$. We verified that the energy loss into QBSs agrees well with the one obtained in the hydrogenic approximation in~\cite{Tomaselli:2023ysb}, with only minor corrections due to our use of a toy-model potential. 
As shown in the figure, energy loss due to QNM emission is generally subdominant, but it becomes comparable to the loss into QBSs when the particle passes close to the BH and for larger values of $\alpha$. This is because QNM excitation is more efficient when the particle’s orbital frequency is high enough to bridge the gap between the QBS and QNM frequencies. Such conditions naturally arise for small periapsis, where the instantaneous frequency of the orbit is higher. Furthermore, as shown in Fig.~\ref{fig:overlap}, the spatial overlap between QNMs and QBSs is significantly enhanced when the particle is closer to the BH. Increasing $\alpha$ further amplifies the excitation by reducing the frequency gap between the modes and increasing their overlap. 

\begin{figure}[t]
\centering
\includegraphics[width=\linewidth]{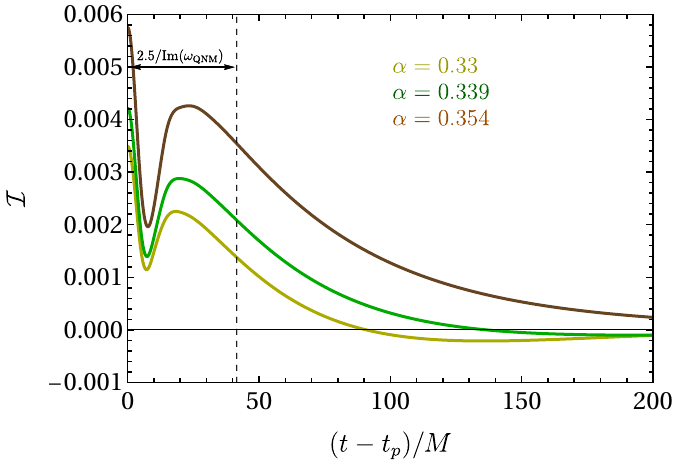}
\caption{Integrand of Eq.~\eqref{eq:integral} as a function the periapsis crossing time for different $\alpha$. As $\alpha$ increases, the integrand decays slower than an e-folding timescale (dashed line).}
\label{fig:secular}
\end{figure}

In Fig.~\ref{fig:Elost}, we stopped at a value of $\alpha=0.339$. Increasing $\alpha$ even further would likely enhance QNM emission, and possibly make it dominant over the QB one for small periastrii. However, increasing $\alpha$ further pushes us outside the regime of validity of our formalism. In solving the Schrödinger-like equation with a simple time-integration, we neglected the imaginary part of the QNM frequencies in the bilinear form. This contribution corresponds to a secular term, varying on an e-folding timescale, which would require a two-timescale expansion to be consistently included. Our approximation is justified because the integrand is strongly localized around periapsis passage, while it decays rapidly and oscillates away from the peak, so that the excitation occurs in a sudden, sharply peaked burst around the peak rather than over an extended timescale
(see~\cite{Redondo-Yuste:2023ipg} for a similar approximation in the gravitational case). When the QNM and QBS frequencies are sufficiently separated, this decay occurs efficiently within a few e-folding timescale, and the secular effects remain negligible. As $\alpha$ increases, however, the two frequencies approach each other, slowing down the decay of the integrand. Physically, this reflects the fact that the particle can bridge the small frequency gap not only near periapsis, but also at times when its instantaneous orbital frequency is lower. In this regime, the secular term cannot be neglected, and a more refined treatment would be required. Fig.~\ref{fig:secular} shows the integrand $\mathcal{I}$ appearing in eq.~\eqref{eq:integral} for $r_p=3M$ as a function of the periastron crossing time for different $\alpha$. As shown in the figure, the integrand decay is slower when $\alpha$ increases. Taking a reference timescale of $2.5$ e-folds, over which the exponential varies by a factor of approximately $10$, we find that at $\alpha = 0.33$ the integral accounts for about $90\%$ of the total, while at $\alpha = 0.354$ it amounts to only about $50\%$, preventing the computation from being extended to higher $\alpha$ values (for this reason, $\alpha=0.354$ is not shown in Fig.~\ref{fig:Elost}).

Our results are obtained under a number of simplifying assumptions -- a Schwarzschild background, a smoothened Newtonian potential, and neglecting secular effects -- and should therefore be regarded as order-of-magnitude estimates. Nevertheless, they clearly indicate that QNMs can be efficiently excited in burst-like events by unbound particles with sufficiently small periapsis, potentially dominating over QBS energy loss at high $\alpha$, and thus represent a non-negligible channel of energy loss during capture processes. This is consistent with previous findings in the gravitational case, where it was found that gravitational QNMs can be excited appreciably by particles passing very close to the light-ring ~\cite{Thornburg:2019ukt, Thornburg:2016msc, OSullivan:2014ywd, 10.1143/PTP.72.494}. 
 \\
Finally, note that while in this paper we considered circular (resonant) and unbound (non-resonant) orbits, one could also study bound, eccentric orbits. Depending on their eccentricity, transitions induced by such orbits can interpolate between the two cases we studied. For small eccentricity, the orbit spends much of its time near the same radius, allowing resonances to build adiabatically, while for large eccentricity, the orbit rapidly samples different radii, suppressing resonant effects~\cite{Tomaselli:2023ysb}. Thus, the phenomenology of eccentric orbits lies naturally between the resonant and non-resonant limits analyzed here.

\section{Conclusions}\label{sec:conclusions}

In this work, we investigated the mathematical properties of QNMs and QBS of massive scalar fields, and estimated QB–to-QN transitions in boson clouds. We first established, both analytically and numerically, the orthogonality between the two mode families with respect to the recently introduced bilinear form for massive scalar fields \cite{Cannizzaro:2023jle}. We also showed how the bilinear form is related to QNM and QBS excitation coefficients in the Laplace formalism, by adapting methods from leaky-wave optics. Finally, we demonstrated that transitions between QBSs and QNMs can occur in the presence of an external potential, and that although such transitions are generally suppressed when the potential is supported far from the BH, they become appreciable for potentials concentrated near it. We analyzed in detail the case of a tidal perturbation generated by a binary companion. We showed that QNM emission is suppressed in resonant systems, such as circular binaries, while it becomes significant in non-resonant capture processes. In the latter case, we employed a few approximations. First, we used a toy-model potential that mimics the Newtonian interaction, and second, we restricted our analysis to the region of parameter space where secular effects are negligible.

A natural extension of our work would be to refine this setup by employing a fully relativistic description of the tidal potential sourced by a fly-by particle. This could be obtained by generalizing the perturbative framework of Ref.~\cite{Brito:2023pyl} to unbound orbits.
In this work, we also restricted our attention to fundamental $\ell=1$ QNM transitions. A more complete analysis should incorporate higher multipoles and overtones, which are expected to enhance the excitation in relevant regimes. 
Another natural extension would be to include the spin of the primary BH, as studies in the gravitational case have shown that the QNM excitation amplitude is strongly dependent on the BH spin~\cite{Thornburg:2019ukt, Thornburg:2016msc, OSullivan:2014ywd, 10.1143/PTP.72.494}. 

Another interesting direction would be to consider other astrophysical perturbations as sources of the QN-QB transition: for instance, accretion-disk turbulence has been shown to excite gravitational QNMs around Kerr BHs~\cite{Yuan:2025fde}; if a boson cloud were present, similar processes could also induce scalar QNM excitation and potentially deplete the cloud. Moreover, binary extreme mass-ratio inspirals (b-EMRIs) evolving inside boson clouds provide a promising laboratory: the high internal orbital frequency of these systems could efficiently excite high-frequency scalar QNMs, in close analogy with the gravitational case~\cite{Santos:2025ass}, and in particular high multipoles that display strong overlap with the cloud wavefunction.

Overall, our work motivates further studies of boson clouds in astrophysical environments within a fully relativistic framework.
We hope to report on these interesting problems in the future.

\vspace{1cm}

\begin{acknowledgments}
We thank Thomas Spieksma, Conor Dyson and Jaime Redondo-Yuste for useful conversations. We acknowledge financial support by the VILLUM Foundation (grant no. VIL37766) and the DNRF Chair program (grant no. DNRF162) by the Danish National Research Foundation. 
E.C. acknowledges financial support
provided under the European Union’s H2020 ERC Advanced Grant “Black holes: gravitational engines of discovery” grant agreement no. Gravitas–101052587. Views
and opinions expressed are however those of the author
only and do not necessarily reflect those of the European Union or the European Research Council. Neither the European Union nor the granting authority can be held responsible for them.
R.B. acknowledge financial support provided by FCT – Fundação para a Ciência e a Tecnologia, I.P., through the ERC-Portugal program Project ``GravNewFields''. We also thank the Fundação para a Ciência e Tecnologia (FCT), Portugal, for the financial support to the Center for Astrophysics and Gravitation (CENTRA/IST/ULisboa) through grant No. UID/PRR/00099/2025 (https://doi.org/10.54499/UID/PRR/00099/2025) and grant No. UID/00099/2025 (https://doi.org/10.54499/UID/00099/2025).
L.S.~acknowledges support from the UKRI Horizon guarantee funding (project no. EP/Y023706/1). L.S.~is also supported by a University of Nottingham Anne McLaren Fellowship. S.R.G. is supported by a UKRI Future Leaders Fellowship (grant number MR/Y018060/1).
\end{acknowledgments}





\bibliography{ref}

@ARTICLE{2024MNRAS.534...39B,
       author = {{Brown}, Joshua J. and {Ogilvie}, Gordon I.},
        title = "{Horseshoes and spiral waves: capturing the 3D flow induced by a low-mass planet analytically}",
      journal = {"Monthly Notices of the Royal Astronomical Society"},
         year = 2024,
        month = oct,
       volume = {534},
       number = {1},
        pages = {39-55},
          doi = {10.1093/mnras/stae2060},
archivePrefix = {arXiv},
       eprint = {2409.02687},
 primaryClass = {astro-ph.EP},
       adsurl = {https://ui.adsabs.harvard.edu/abs/2024MNRAS.534...39B}
}

@ARTICLE{2020ApJ...892...65M,
       author = {{Miranda}, Ryan and {Rafikov}, Roman R.},
        title = "{Planet-Disk Interaction in Disks with Cooling: Basic Theory}",
      journal = {\apj},
         year = 2020,
        month = mar,
       volume = {892},
       number = {1},
          eid = {65},
        pages = {65},
          doi = {10.3847/1538-4357/ab791a},
archivePrefix = {arXiv},
       eprint = {1911.01428},
 primaryClass = {astro-ph.EP},
       adsurl = {https://ui.adsabs.harvard.edu/abs/2020ApJ...892...65M}
}

@article{Li:2025ffh,
    author = "Li, Dongjun and Weller, Colin and Bourg, Patrick and LaHaye, Michael and Yunes, Nicol{\'a}s and Yang, Huan",
    title = "{Extreme mass-ratio inspiral within an ultralight scalar cloud: Scalar radiation}",
    eprint = "2507.02045",
    archivePrefix = "arXiv",
    primaryClass = "gr-qc",
    doi = "10.1103/7l9s-g21j",
    journal = "Phys. Rev. D",
    volume = "112",
    number = "8",
    pages = "084057",
    year = "2025"
}

@article{DellaMonica:2025zby,
    author = "Della Monica, Riccardo and Brito, Richard",
    title = "{Detectability of gravitational atoms in black hole binaries with the Einstein Telescope}",
    eprint = "2503.23419",
    archivePrefix = "arXiv",
    primaryClass = "gr-qc",
    reportNumber = "ET-0056A-25",
    doi = "10.1103/h7ld-vv9p",
    journal = "Phys. Rev. D",
    volume = "112",
    number = "2",
    pages = "024074",
    year = "2025"
}

@article{Percival:2020skc,
    author = "Percival, Jake and Dolan, Sam R.",
    title = "{Quasinormal modes of massive vector fields on the Kerr spacetime}",
    eprint = "2008.10621",
    archivePrefix = "arXiv",
    primaryClass = "gr-qc",
    doi = "10.1103/PhysRevD.102.104055",
    journal = "Phys. Rev. D",
    volume = "102",
    number = "10",
    pages = "104055",
    year = "2020"
}

@article{Santos:2025ass,
    author = "Santos, Jo{\~a}o S. and Cardoso, Vitor and Nat{\'a}rio, Jos{\'e} and van de Meent, Maarten",
    title = "{Gravitational Waves from Binary Extreme Mass Ratio Inspirals: Doppler Shift and Beaming, Resonant Excitation, Helicity Oscillations, and Self-Lensing}",
    eprint = "2506.14868",
    archivePrefix = "arXiv",
    primaryClass = "gr-qc",
    doi = "10.1103/qq3m-6phh",
    journal = "Phys. Rev. Lett.",
    volume = "135",
    number = "21",
    pages = "211402",
    year = "2025"
}

@article{Redondo-Yuste:2023ipg,
    author = "Redondo-Yuste, Jaime and Pere{\~n}iguez, David and Cardoso, Vitor",
    title = "{Ringdown of a dynamical spacetime}",
    eprint = "2312.04633",
    archivePrefix = "arXiv",
    primaryClass = "gr-qc",
    doi = "10.1103/PhysRevD.109.044048",
    journal = "Phys. Rev. D",
    volume = "109",
    number = "4",
    pages = "044048",
    year = "2024"
}

@misc{GrAB,
  author       = {},
  title        = {},
  year         = {},
  howpublished = {\url{https://github.com/thomasspieksma/GrAB}},
}

@article{Yuan:2025fde,
    author = "Yuan, Chen and Cardoso, Vitor and Duque, Francisco and Younsi, Ziri",
    title = "{Gravitational waves from accretion disks: Turbulence, mode excitation, and prospects for future detectors}",
    eprint = "2502.07871",
    archivePrefix = "arXiv",
    primaryClass = "gr-qc",
    doi = "10.1103/PhysRevD.111.063048",
    journal = "Phys. Rev. D",
    volume = "111",
    number = "6",
    pages = "063048",
    year = "2025"
}

@article{Decanini:2014bwa,
    author = "D\'ecanini, Yves and Folacci, Antoine and Ould El Hadj, Mohamed",
    title = "{Resonant excitation of black holes by massive bosonic fields and giant ringings}",
    eprint = "1402.2481",
    archivePrefix = "arXiv",
    primaryClass = "gr-qc",
    doi = "10.1103/PhysRevD.89.084066",
    journal = "Phys. Rev. D",
    volume = "89",
    number = "8",
    pages = "084066",
    year = "2014"
}

@article{Cavalcante:2024kmy,
    author = "Cavalcante, Jo\~ao Paulo and Richartz, Maur\'\i{}cio and da Cunha, Bruno Carneiro",
    title = "{Massive scalar perturbations in Kerr black holes: Near extremal analysis}",
    eprint = "2408.13964",
    archivePrefix = "arXiv",
    primaryClass = "gr-qc",
    doi = "10.1103/PhysRevD.110.124064",
    journal = "Phys. Rev. D",
    volume = "110",
    number = "12",
    pages = "124064",
    year = "2024"
}

@article{10.1121/1.4973313,
    author = {Gallezot, Matthieu and Treyssède, Fabien and Laguerre, Laurent},
    title = {Contribution of leaky modes in the modal analysis of unbounded problems with perfectly matched layers},
    journal = {The Journal of the Acoustical Society of America},
    volume = {141},
    number = {1},
    pages = {EL16-EL21},
    year = {2017},
    month = {01},
    issn = {0001-4966},
    doi = {10.1121/1.4973313},
    url = {https://doi.org/10.1121/1.4973313},
}

@article{Feldbrugge:2017kzv,
    author = "Feldbrugge, Job and Lehners, Jean-Luc and Turok, Neil",
    title = "{Lorentzian Quantum Cosmology}",
    eprint = "1703.02076",
    archivePrefix = "arXiv",
    primaryClass = "hep-th",
    doi = "10.1103/PhysRevD.95.103508",
    journal = "Phys. Rev. D",
    volume = "95",
    number = "10",
    pages = "103508",
    year = "2017"
}

@book{Leaky1,
    author = "Tamir, T and Oliner, A",
    title = "{Guided Complex Waves: Part I. Fields at an Interface}",
    eprint = "",
    archivePrefix = "v",
    primaryClass = "",
    doi = "",
    journal = "",
    volume = "",
    number = "",
    pages = "",
    year = "1962"
}

@article{Decanini:2015yba,
    author = "Decanini, Yves and Folacci, Antoine and Ould El Hadj, Mohamed",
    title = "{Waveforms produced by a scalar point particle plunging into a Schwarzschild black hole: Excitation of quasinormal modes and quasibound states}",
    eprint = "1506.09133",
    archivePrefix = "arXiv",
    primaryClass = "gr-qc",
    doi = "10.1103/PhysRevD.92.024057",
    journal = "Phys. Rev. D",
    volume = "92",
    number = "2",
    pages = "024057",
    year = "2015"
}

@article{Ma:2024qcv,
    author = "Ma, Sizheng and Yang, Huan",
    title = "{Excitation of quadratic quasinormal modes for Kerr black holes}",
    eprint = "2401.15516",
    archivePrefix = "arXiv",
    primaryClass = "gr-qc",
    doi = "10.1103/PhysRevD.109.104070",
    journal = "Phys. Rev. D",
    volume = "109",
    number = "10",
    pages = "104070",
    year = "2024"
}

@article{10.1143/PTP.72.494,
    author = {Kojima, Yasufumi and Nakamura, Takashi},
    title = {Gravitational Radiation from a Particle Scattered by a Kerr Black Hole},
    journal = {Progress of Theoretical Physics},
    volume = {72},
    number = {3},
    pages = {494-504},
    year = {1984},
    month = {09},
    issn = {0033-068X},
    doi = {10.1143/PTP.72.494},
    url = {https://doi.org/10.1143/PTP.72.494},
    eprint = {https://academic.oup.com/ptp/article-pdf/72/3/494/5382232/72-3-494.pdf},
}

@article{Thornburg:2016msc,
    author = "Thornburg, Jonathan and Wardell, Barry",
    title = "{Scalar self-force for highly eccentric equatorial orbits in Kerr spacetime}",
    eprint = "1610.09319",
    archivePrefix = "arXiv",
    primaryClass = "gr-qc",
    doi = "10.1103/PhysRevD.95.084043",
    journal = "Phys. Rev. D",
    volume = "95",
    number = "8",
    pages = "084043",
    year = "2017"
}

@article{Thornburg:2019ukt,
    author = "Thornburg, Jonathan and Wardell, Barry and van de Meent, Maarten",
    title = "{Excitation of Kerr quasinormal modes in extreme--mass-ratio inspirals}",
    eprint = "1906.06791",
    archivePrefix = "arXiv",
    primaryClass = "gr-qc",
    doi = "10.1103/PhysRevResearch.2.013365",
    journal = "Phys. Rev. Res.",
    volume = "2",
    number = "1",
    pages = "013365",
    year = "2020"
}

@article{Cannizzaro:2023jle,
    author = "Cannizzaro, Enrico and Sberna, Laura and Green, Stephen R. and Hollands, Stefan",
    title = "{Relativistic Perturbation Theory for Black-Hole Boson Clouds}",
    eprint = "2309.10021",
    archivePrefix = "arXiv",
    primaryClass = "gr-qc",
    doi = "10.1103/PhysRevLett.132.051401",
    journal = "Phys. Rev. Lett.",
    volume = "132",
    number = "5",
    pages = "051401",
    year = "2024"
}

@article{Dyson:2025dlj,
    author = "Dyson, Conor and Spieksma, Thomas F. M. and Brito, Richard and van de Meent, Maarten and Dolan, Sam",
    title = "{Environmental Effects in Extreme-Mass-Ratio Inspirals: Perturbations to the Environment in Kerr Spacetimes}",
    eprint = "2501.09806",
    archivePrefix = "arXiv",
    primaryClass = "gr-qc",
    doi = "10.1103/PhysRevLett.134.211403",
    journal = "Phys. Rev. Lett.",
    volume = "134",
    number = "21",
    pages = "211403",
    year = "2025"
}

@article{Duque:2023seg,
    author = "Duque, Francisco and Macedo, Caio F. B. and Vicente, Rodrigo and Cardoso, Vitor",
    title = "{Extreme-Mass-Ratio Inspirals in Ultralight Dark Matter}",
    eprint = "2312.06767",
    archivePrefix = "arXiv",
    primaryClass = "gr-qc",
    doi = "10.1103/PhysRevLett.133.121404",
    journal = "Phys. Rev. Lett.",
    volume = "133",
    number = "12",
    pages = "121404",
    year = "2024"
}

@article{Tomaselli:2024bdd,
    author = "Tomaselli, Giovanni Maria and Spieksma, Thomas F. M. and Bertone, Gianfranco",
    title = "{Resonant history of gravitational atoms in black hole binaries}",
    eprint = "2403.03147",
    archivePrefix = "arXiv",
    primaryClass = "gr-qc",
    doi = "10.1103/PhysRevD.110.064048",
    journal = "Phys. Rev. D",
    volume = "110",
    number = "6",
    pages = "064048",
    year = "2024"
}

@article{Cole:2022yzw,
    author = "Cole, Philippa S. and Bertone, Gianfranco and Coogan, Adam and Gaggero, Daniele and Karydas, Theophanes and Kavanagh, Bradley J. and Spieksma, Thomas F. M. and Tomaselli, Giovanni Maria",
    title = "{Distinguishing environmental effects on binary black hole gravitational waveforms}",
    eprint = "2211.01362",
    archivePrefix = "arXiv",
    primaryClass = "gr-qc",
    doi = "10.1038/s41550-023-01990-2",
    journal = "Nature Astron.",
    volume = "7",
    number = "8",
    pages = "943--950",
    year = "2023"
}

@article{Zhang:2019eid,
    author = "Zhang, Jun and Yang, Huan",
    title = "{Dynamic Signatures of Black Hole Binaries with Superradiant Clouds}",
    eprint = "1907.13582",
    archivePrefix = "arXiv",
    primaryClass = "gr-qc",
    reportNumber = "Imperial/TP/2019/JZ/02",
    doi = "10.1103/PhysRevD.101.043020",
    journal = "Phys. Rev. D",
    volume = "101",
    number = "4",
    pages = "043020",
    year = "2020"
}

@ARTICLE{AmaroSeoane2017,
       author = {{Amaro-Seoane}, Pau and {Audley}, Heather and {Babak}, Stanislav and {Baker}, John and {Barausse}, Enrico and {Bender}, Peter and {Berti}, Emanuele and {Binetruy}, Pierre and {Born}, Michael and {Bortoluzzi}, Daniele and {Camp}, Jordan and {Caprini}, Chiara and {Cardoso}, Vitor and {Colpi}, Monica and {Conklin}, John and {Cornish}, Neil and {Cutler}, Curt and {Danzmann}, Karsten and {Dolesi}, Rita and {Ferraioli}, Luigi and {Ferroni}, Valerio and {Fitzsimons}, Ewan and {Gair}, Jonathan and {Gesa Bote}, Lluis and {Giardini}, Domenico and {Gibert}, Ferran and {Grimani}, Catia and {Halloin}, Hubert and {Heinzel}, Gerhard and {Hertog}, Thomas and {Hewitson}, Martin and {Holley-Bockelmann}, Kelly and {Hollington}, Daniel and {Hueller}, Mauro and {Inchauspe}, Henri and {Jetzer}, Philippe and {Karnesis}, Nikos and {Killow}, Christian and {Klein}, Antoine and {Klipstein}, Bill and {Korsakova}, Natalia and {Larson}, Shane L and {Livas}, Jeffrey and {Lloro}, Ivan and {Man}, Nary and {Mance}, Davor and {Martino}, Joseph and {Mateos}, Ignacio and {McKenzie}, Kirk and {McWilliams}, Sean T and {Miller}, Cole and {Mueller}, Guido and {Nardini}, Germano and {Nelemans}, Gijs and {Nofrarias}, Miquel and {Petiteau}, Antoine and {Pivato}, Paolo and {Plagnol}, Eric and {Porter}, Ed and {Reiche}, Jens and {Robertson}, David and {Robertson}, Norna and {Rossi}, Elena and {Russano}, Giuliana and {Schutz}, Bernard and {Sesana}, Alberto and {Shoemaker}, David and {Slutsky}, Jacob and {Sopuerta}, Carlos F. and {Sumner}, Tim and {Tamanini}, Nicola and {Thorpe}, Ira and {Troebs}, Michael and {Vallisneri}, Michele and {Vecchio}, Alberto and {Vetrugno}, Daniele and {Vitale}, Stefano and {Volonteri}, Marta and {Wanner}, Gudrun and {Ward}, Harry and {Wass}, Peter and {Weber}, William and {Ziemer}, John and {Zweifel}, Peter},
        title = "{Laser Interferometer Space Antenna}",
      journal = {arXiv e-prints},
         year = 2017,
        month = feb,
          eid = {arXiv:1702.00786},
        pages = {arXiv:1702.00786},
          doi = {10.48550/arXiv.1702.00786},
archivePrefix = {arXiv},
       eprint = {1702.00786},
 primaryClass = {astro-ph.IM},
       adsurl = {https://ui.adsabs.harvard.edu/abs/2017arXiv170200786A}
}

@article{Amaro-Seoane:2012lgq,
    author = "Amaro-Seoane, Pau",
    title = "{Relativistic dynamics and extreme mass ratio inspirals}",
    eprint = "1205.5240",
    archivePrefix = "arXiv",
    primaryClass = "astro-ph.CO",
    doi = "10.1007/s41114-018-0013-8",
    journal = "Living Rev. Rel.",
    volume = "21",
    number = "1",
    pages = "4",
    year = "2018"
}

@article{Tomaselli:2023ysb,
    author = "Tomaselli, Giovanni Maria and Spieksma, Thomas F. M. and Bertone, Gianfranco",
    title = "{Dynamical friction in gravitational atoms}",
    eprint = "2305.15460",
    archivePrefix = "arXiv",
    primaryClass = "gr-qc",
    doi = "10.1088/1475-7516/2023/07/070",
    journal = "JCAP",
    volume = "07",
    pages = "070",
    year = "2023"
}

@article{Baumann:2019ztm,
    author = "Baumann, Daniel and Chia, Horng Sheng and Porto, Rafael A. and Stout, John",
    title = "{Gravitational Collider Physics}",
    eprint = "1912.04932",
    archivePrefix = "arXiv",
    primaryClass = "gr-qc",
    reportNumber = "DESY-19-221, DESY 19-221",
    doi = "10.1103/PhysRevD.101.083019",
    journal = "Phys. Rev. D",
    volume = "101",
    number = "8",
    pages = "083019",
    year = "2020"
}

@article{Leaver:1985ax,
    author = "Leaver, E. W.",
    title = "{An Analytic representation for the quasi normal modes of Kerr black holes}",
    doi = "10.1098/rspa.1985.0119",
    journal = "Proc. Roy. Soc. Lond. A",
    volume = "402",
    pages = "285--298",
    year = "1985"
}

@article{Shahar1998,
  title = {Late-time tails in gravitational collapse of a self-interacting (massive) scalar-field and decay of a self-interacting scalar hair},
  author = {Hod, Shahar and Piran, Tsvi},
  journal = {Phys. Rev. D},
  volume = {58},
  issue = {4},
  pages = {044018},
  numpages = {6},
  year = {1998},
  month = {Jul},
  publisher = {American Physical Society},
  doi = {10.1103/PhysRevD.58.044018},
  url = {https://link.aps.org/doi/10.1103/PhysRevD.58.044018}
}

@article{Brito:2023pyl,
    author = "Brito, Richard and Shah, Shreya",
    title = "{Extreme mass-ratio inspirals into black holes surrounded by scalar clouds}",
    eprint = "2307.16093",
    archivePrefix = "arXiv",
    primaryClass = "gr-qc",
    doi = "10.1103/PhysRevD.108.084019",
    journal = "Phys. Rev. D",
    volume = "108",
    number = "8",
    pages = "084019",
    year = "2023",
    note = "[Erratum: Phys.Rev.D 110, 109902 (2024)]"
}

@article{Berti:2006wq,
    author = "Berti, Emanuele and Cardoso, Vitor",
    title = "{Quasinormal ringing of Kerr black holes. I. The Excitation factors}",
    eprint = "gr-qc/0605118",
    archivePrefix = "arXiv",
    doi = "10.1103/PhysRevD.74.104020",
    journal = "Phys. Rev. D",
    volume = "74",
    pages = "104020",
    year = "2006"
}

@article{Leaver:1986gd,
    author = "Leaver, Edward W.",
    title = "{Spectral decomposition of the perturbation response of the Schwarzschild geometry}",
    doi = "10.1103/PhysRevD.34.384",
    journal = "Phys. Rev. D",
    volume = "34",
    pages = "384--408",
    year = "1986"
}

@article{Koyama:2001ee,
    author = "Koyama, Hiroko and Tomimatsu, Akira",
    title = "{Asymptotic tails of massive scalar fields in Schwarzschild background}",
    eprint = "gr-qc/0103086",
    archivePrefix = "arXiv",
    doi = "10.1103/PhysRevD.64.044014",
    journal = "Phys. Rev. D",
    volume = "64",
    pages = "044014",
    year = "2001"
}

@article{Koyama:2001qw,
    author = "Koyama, Hiroko and Tomimatsu, Akira",
    title = "{Slowly decaying tails of massive scalar fields in spherically symmetric space-times}",
    eprint = "gr-qc/0112075",
    archivePrefix = "arXiv",
    doi = "10.1103/PhysRevD.65.084031",
    journal = "Phys. Rev. D",
    volume = "65",
    pages = "084031",
    year = "2002"
}

@article{PhysRevD.51.353,
  title = {Excitation of Schwarzschild black-hole quasinormal modes},
  author = {Andersson, Nils},
  journal = {Phys. Rev. D},
  volume = {51},
  issue = {2},
  pages = {353--363},
  numpages = {0},
  year = {1995},
  month = {Jan},
  publisher = {American Physical Society},
  doi = {10.1103/PhysRevD.51.353},
  url = {https://link.aps.org/doi/10.1103/PhysRevD.51.353}
}

@article{Nollert:1992ifk,
    author = "Nollert, Hans-Peter and Schmidt, Bernd G.",
    title = "{Quasinormal modes of Schwarzschild black holes: Defined and calculated via Laplace transformation}",
    doi = "10.1103/PhysRevD.45.2617",
    journal = "Phys. Rev. D",
    volume = "45",
    number = "8",
    pages = "2617",
    year = "1992"
}

@article{Green:2022htq,
    author = "Green, Stephen R. and Hollands, Stefan and Sberna, Laura and Toomani, Vahid and Zimmerman, Peter",
    title = "{Conserved currents for a Kerr black hole and orthogonality of quasinormal modes}",
    eprint = "2210.15935",
    archivePrefix = "arXiv",
    primaryClass = "gr-qc",
    doi = "10.1103/PhysRevD.107.064030",
    journal = "Phys. Rev. D",
    volume = "107",
    number = "6",
    pages = "064030",
    year = "2023"
}

@article{Dolan:2007mj,
      author         = "Dolan, Sam R.",
      title          = "{Instability of the massive Klein-Gordon field on the
                        Kerr spacetime}",
      journal        = "Phys.Rev.",
      volume         = "D76",
      pages          = "084001",
      doi            = "10.1103/PhysRevD.76.084001",
      year           = "2007",
      eprint         = "0705.2880",
      archivePrefix  = "arXiv",
      primaryClass   = "gr-qc",
      SLACcitation   = "%%CITATION = ARXIV:0705.2880;%%",
}

@article{Glampedakis:2001js,
    author = "Glampedakis, Kostas and Andersson, Nils",
    title = "{Late time dynamics of rapidly rotating black holes}",
    eprint = "gr-qc/0103054",
    archivePrefix = "arXiv",
    doi = "10.1103/PhysRevD.64.104021",
    journal = "Phys. Rev. D",
    volume = "64",
    pages = "104021",
    year = "2001"
}

@article{Arvanitaki:2010sy,
      author         = "Arvanitaki, Asimina and Dubovsky, Sergei",
      title          = "{Exploring the String Axiverse with Precision Black Hole
                        Physics}",
      journal        = "Phys.Rev.",
      volume         = "D83",
      pages          = "044026",
      doi            = "10.1103/PhysRevD.83.044026",
      year           = "2011",
      eprint         = "1004.3558",
      archivePrefix  = "arXiv",
      primaryClass   = "hep-th",
      SLACcitation   = "%%CITATION = ARXIV:1004.3558;%%",
}

@article{OSullivan:2014ywd,
      author         = "O'Sullivan, Stephen and Hughes, Scott A.",
      title          = "{Strong-field tidal distortions of rotating black holes:
                        Formalism and results for circular, equatorial orbits}",
      journal        = "Phys. Rev.",
      volume         = "D90",
      year           = "2014",
      number         = "12",
      pages          = "124039",
      doi            = "10.1103/PhysRevD.91.109901, 10.1103/PhysRevD.90.124039",
      note           = "[Erratum: Phys. Rev.D91,no.10,109901(2015)]",
      eprint         = "1407.6983",
      archivePrefix  = "arXiv",
      primaryClass   = "gr-qc",
      SLACcitation   = "%%CITATION = ARXIV:1407.6983;%%"
}

@article{Berti:2025hly,
    author = "Berti, Emanuele and others",
    title = "{Black hole spectroscopy: from theory to experiment}",
    eprint = "2505.23895",
    archivePrefix = "arXiv",
    primaryClass = "gr-qc",
    month = "5",
    year = "2025",
    journal = ""
}

@article{Kocsis:2011dr,
    author = "Kocsis, Bence and Yunes, Nicolas and Loeb, Abraham",
    title = "{Observable Signatures of EMRI Black Hole Binaries Embedded in Thin Accretion Disks}",
    eprint = "1104.2322",
    archivePrefix = "arXiv",
    primaryClass = "astro-ph.GA",
    doi = "10.1103/PhysRevD.86.049907",
    journal = "Phys. Rev. D",
    volume = "84",
    pages = "024032",
    year = "2011"
}

@article{Yunes:2011ws,
    author = "Yunes, Nicolas and Kocsis, Bence and Loeb, Abraham and Haiman, Zoltan",
    title = "{Imprint of Accretion Disk-Induced Migration on Gravitational Waves from Extreme Mass Ratio Inspirals}",
    eprint = "1103.4609",
    archivePrefix = "arXiv",
    primaryClass = "astro-ph.CO",
    doi = "10.1103/PhysRevLett.107.171103",
    journal = "Phys. Rev. Lett.",
    volume = "107",
    pages = "171103",
    year = "2011"
}

@article{Speri:2022upm,
    author = "Speri, Lorenzo and Antonelli, Andrea and Sberna, Laura and Babak, Stanislav and Barausse, Enrico and Gair, Jonathan R. and Katz, Michael L.",
    title = "{Probing Accretion Physics with Gravitational Waves}",
    eprint = "2207.10086",
    archivePrefix = "arXiv",
    primaryClass = "gr-qc",
    doi = "10.1103/PhysRevX.13.021035",
    journal = "Phys. Rev. X",
    volume = "13",
    number = "2",
    pages = "021035",
    year = "2023"
}

@book{Brito:2015oca,
    author = "Brito, Richard and Cardoso, Vitor and Pani, Paolo",
    title = "{Superradiance}: {New Frontiers in Black Hole
Physics}",
    eprint = "1501.06570",
    archivePrefix = "arXiv",
    primaryClass = "gr-qc",
    doi = "10.1007/978-3-319-19000-6",
    isbn = "978-3-319-18999-4, 978-3-319-19000-6",
    publisher = "Springer",
    volume = "906",
    year = "2015"
}

@article{Arvanitaki:2009fg,
      author         = "Arvanitaki, Asimina and Dimopoulos, Savas and Dubovsky,
                        Sergei and Kaloper, Nemanja and March-Russell, John",
      title          = "{String Axiverse}",
      journal        = "Phys. Rev.",
      volume         = "D81",
      year           = "2010",
      pages          = "123530",
      doi            = "10.1103/PhysRevD.81.123530",
      eprint         = "0905.4720",
      archivePrefix  = "arXiv",
      primaryClass   = "hep-th",
      SLACcitation   = "%%CITATION = ARXIV:0905.4720;%%"
}

@article{Baumann:2018vus,
    author = "Baumann, Daniel and Chia, Horng Sheng and Porto, Rafael A.",
    title = "{Probing Ultralight Bosons with Binary Black Holes}",
    eprint = "1804.03208",
    archivePrefix = "arXiv",
    primaryClass = "gr-qc",
    reportNumber = "DESY-18-060, DESY 18-060",
    doi = "10.1103/PhysRevD.99.044001",
    journal = "Phys. Rev. D",
    volume = "99",
    number = "4",
    pages = "044001",
    year = "2019"
}

@article{Baumann:2021fkf,
    author = "Baumann, Daniel and Bertone, Gianfranco and Stout, John and Tomaselli, Giovanni Maria",
    title = "{Ionization of gravitational atoms}",
    eprint = "2112.14777",
    archivePrefix = "arXiv",
    primaryClass = "gr-qc",
    doi = "10.1103/PhysRevD.105.115036",
    journal = "Phys. Rev. D",
    volume = "105",
    number = "11",
    pages = "115036",
    year = "2022"
}

@article{Baumann:2022pkl,
    author = "Baumann, Daniel and Bertone, Gianfranco and Stout, John and Tomaselli, Giovanni Maria",
    title = "{Sharp Signals of Boson Clouds in Black Hole Binary Inspirals}",
    eprint = "2206.01212",
    archivePrefix = "arXiv",
    primaryClass = "gr-qc",
    doi = "10.1103/PhysRevLett.128.221102",
    journal = "Phys. Rev. Lett.",
    volume = "128",
    number = "22",
    pages = "221102",
    year = "2022"
}

\end{document}